\documentclass[10pt,a4paper]{article}
\usepackage{t1enc}
\usepackage[latin1]{inputenc}
\usepackage[german,english]{babel}
\usepackage{amsfonts}

\newtheorem{defi}{Definition}[section]
\newtheorem{prop}[defi]{Proposition}

\newtheorem{coro}[defi]{Corollary}

\begin{document}

\title{{\bf Non-associative public-key cryptography}}
\author{Arkadius Kalka}
\date{} 
\maketitle

\begin{abstract} We introduce a generalized Anshel-Anshel-Goldfeld (AAG) key establishment protocol (KEP) for magmas. 
This leads to the foundation of non-associative public-key cryptography (PKC), generalizing the concept of non-commutative PKC. 
We show that left selfdistributive systems appear in a natural special case of a generalized AAG-KEP for magmas, 
and we propose, among others instances, concrete realizations using $f$-conjugacy in groups and shifted conjugacy in braid groups.
We discuss the advantages of our schemes compared with the classical AAG-KEP based on conjugacy in braid groups.
\end{abstract} 

\section{Introduction}

Currently public key cryptography still relies mainly on a few number-theoretic problems, namely integer factorization \cite{RSA78} and the computation of
discrete logarithms in $\mathbb{Z}_p^{\times }$ and over elliptic curves.
The systems based on these problems remain unbroken. 
Nevertheless, after the advent of quantum computers, systems like RSA \cite{RSA78} and its variants (e.g. \cite{Ra79}), Diffie-Hellman (DH) \cite{DH76}, ElGamal \cite{El85} and ECC \cite{Mi85,Ko87} will be broken easily \cite{Sh97,PZ03}. \\
Under the label Post Quantum Cryptography, there have been several efforts to develop new cryptographic primitives which may also serve for the post quantum computer era. Here we focus on key establishment protocols (KEP's) as cryptographic primitives, because they are the most important and the hardest to construct. Note that, using hash functions, it is easy to build public key encryption schemes from KEP's. \\
One approach became later known as {\it non-commutative cryptography}. Recall that the involved algebraic structures in the number-theoretic systems are
commutative groups and rings. In non-commutative cryptography these are replaced by non-commutative groups and rings, 
and we consider computational problems therein. 
One may say that, roughly, the discrete logarithm problem is replaced by the conjugacy problem and its variants.
After some precursors, in particular \cite{WM85}, non-commutative cryptography was mainly established
in a few seminal papers around the turn of the millenium \cite{AAG99, KL+00, CK+01}. Of particular importance is the ingenious Anshel-Anshel-Goldfeld (AAG)
Commutator KEP which only exists in the non-commutative setting, while the systems in \cite{KL+00, CK+01} may be considered as staightforward non-commutative 
analogues of the classical DH-KEP. \\
Since they admit efficiently computable normal forms and a supposedly hard conjugacy problem, braid groups were explicitly suggested as platform groups for these systems. Nevertheless, explicit specifications of these systems in braid groups as well as most other non-commutative cryptosystems have been broken over the last
decade. This led to some understandable decline of interest in non-commutative cryptography inside the main cryptographic community. 
A revival of non-commutative cryptography may be achieved by means of research in one of the following two directions. \\
The first approach is to stick with the suggested protocols and search for better platform groups. One may even keep braid groups as platforms and search for families of hard instances of the conjugacy problem that can be efficiently generated. Note that the main reason why braid-based cryptosystems have been broken is the fact that "randomly" generated keys turned out to be a very bad choice. This situation is quite typical for public-key cryptography. Consider, for example, the familiar RSA scheme where the keys have to be chosen with care. \\
Another approach is to construct new or generalized non-commutative cryptosystems which are based on other or supposedly harder computational problems.
In this and some subsequent papers we pursue the latter approach. In particular, we broaden the scope of non-commutative cryptography as we go beyond non-commutative, associative binary oparations - we utilize non-associative binary operations, i.e. magmas. Thus, we hope to establish the field
of {\it non-associative public-key cryptography}. In particular, we generalize the AAG-KEP for monoids to a general AAG-KEP for magmas. \\

{\bf Outline.} The paper is structured as follows. In section \ref{AAGKE} we emphasize the important and integrating role of the AAG protocol in non-commutative and commutative cryptography. In particular, we introduce a generalized notion of AAG-KEP for monoids (section \ref{AAGmon}), and we show that not only the AAG commutator KEP for groups \cite{AAG99} (section \ref{AAGgr}), but also the Ko-Lee et al. protocol, the group Diffie-Hellman protocol (section \ref{grDHKE}), and even the classical DH-KEP (section \ref{DHKE}) are special instances of that generalized AAG scheme. Furthermore, we also subsume the 
Sakalauskas, Tvarijonas and Raulynaitis KEP(STR-KEP), a natural hybrid of the classical DH-KEP and the Ko-Lee-KEP, 
as a further instance (section \ref{STRKEP}).  \\
The main innovative part of this paper is contained in the sections \ref{nassKE} and \ref{LD}.
In particular, in section \ref{AAGmagma} we extend the generalized AAG-KEP from monoids to magmas. 
Here finitely generated submonoids are replaced by f.g. submagmas, and Alice and Bob know their secret key submagma elements as products of the generators, including planar rooted binary trees describing the bracket structure of such products.
First examples of instances of the generalized AAG-KEP for magmas are a non-associative KEP based on simultaneous double coset problem and symmetric decomposition problem (see sections \ref{simDCP} and \ref{symDP}).   \\
The most interesting and natural instances of the generalized AAG-KEP for magmas come from left-selfdistributive (LD) systems and their generalizations (section \ref{LD}).  In section \ref{LDmult} we introduce LD- and multi-LD-systems with $f$-conjugacy in groups and shifted conjugacy in braid groups as key examples for LD-operations. The nonassociative AAG $f$-commutator KEP (section \ref{AAGfCommKEP}) and the AAG shifted commutator KEP (section \ref{AAGshCommKEP}) are discussed as major examples.
We note that for these instances we may even drop the simultaneity of the underlying base problems ($f$- and shifted conjugacy problems), because here submagmas generated by one element still have a rich and complicated structure and a hard membership problem.
This implies that these systems are the first KEP's based on the shifted and $f$-conjugacy problem.  \\
In section \ref{Open} we discuss generalizations, like AAG-schemes over non-associative magmas, open problems and further work. \\

{\bf Summary.} The main purpose of this paper is popularize the notion of non-associative cryptography and to provide a general framework for 
non-associative and non-commutative KEP's by utilizing the unifying approach that stems from the general AAG-KEP for magmas.
We argue for the superiority of the non-associative schemes introduced in section \ref{LD} compared to classical non-commutative AAG commutator KEP. \\
Anyway, in our opinion the field of non-commutative cryptography lacked over the last years supply of new innovative cryptosystems.
We hope that non-associative cryptography will contribute to revived interest in non-commutative cryptography.  \\

{\bf Outlook.} Nevertheless, this is not the end, rather the beginning of the story of non-associative cryptography. \\
In the forthcoming paper \cite{KaT12}, by introducing a small asymmetry in the non-associative AAG protocol for magmas, we succeed to construct 
non-associative KEP's for all LD- and multi-LD-systems (in general: sets with distributive operations).
We consider the systems and instances given in \cite{KaT12} as much more practical and interesting than the one given in this paper.
In particular, since these systems work for all LD- and multi-LD-systems, they deploy two further advantages. 
First, we may consider encryption functions using iterated multiplication (in the magma) from the left. 
Therefore, in order to obtain the secret key an attacker has to solve an iterated $f$- or shifted conjugacy problem. 
Second, for a given (partial) multi-LD-system it turns out that even the used operations can be hidden, i.e., they are part of the secret key.  \\

{\bf Historical remarks.} Non-associative structures, in particular quasigroups seem to have a long history in cryptography. For an overview on cryptographic applications of quasigroups and Latin squares, see \cite{Shc09, GS10, Shc12}. In particular, we mention the work of Denes and Keedwell \cite{DK74, DK91, DK92, DK02}. Nevertheless, except for authentication schemes and zero-knowledge protocols, most of these applications are in classical (i.e. symmetric key) cryptography.
The earliest quasigroup-based public-key cryptosystem that we are aware of is due to Koscielny and Mullen \cite{KM99}. \\
Non-associative cryptography that goes beyond quasigroups, in particular, the generalized AAG-KEP for magmas were introduced by the author in his PhD thesis in 2007 \cite{Ka07}.
During a postdoctoral stay at the Bar-Ilan University, hosted by M. Teicher, we had the opportunity to refine and improve our non-associative systems.
In particular, we developed the non-associative KEP's for all distributive systems \cite{KaT12}. Over the last years we also had the opportunity to promote
non-associative cryptography at several conferences, in particular in Dortmund 2007, Hoboken 2009, Montreal 2010, Caen 2011.  \\
Other non-associative cryptosystems that came up during the last few years include \cite{GMK08, MZ12}. \\

{\bf Acknowledgements.} This work is an extension of a part of my PhD thesis. Therefore, I wish to thank my supervisor L. Gerritzen for his kind support, encouragement, constant interest and steadfast patience. In particular, his great interest in non-associatve algebraic structures as well as public key cryptography formed the scientific environment that made me bring these subjects together. \\
I am greatly indebted to P. Dehornoy who introduced an authentication scheme based on his notion of shifted conjugacy \cite{De06}. 
This in the first place inspired me to come up with a KEP based on shifted conjugacy and in the course of this work to invent non-associative cryptography.  \\
I thank M. Teicher who was my host during my postdoctoral studies at Bar-Ilan University, Israel, in 2007-2011. 
For that time period I acknowledge financial support by The Oswald Veblen Fund and by the Minerva Foundation of Germany. \\  
This paper was written up during my stay at the MPIM Bonn, Jan-March 2012, and finished during my postdoctoral stay at UQ, Brisbane.
For the latter I acknowledge support by the Australian Research Council (project DP110101104).\\  
For valuable and stimulating discussions I thank L. Gerritzen, R. Holtkamp and R. Avanzi at Ruhr-University Bochum,
 M. Kreuzer and G. Rosenberger in Dortmund, B. Tsaban, D. Goldfeld, B. Kunyavskij and R. Cohen at BIU,
A. Myasnikov,  A. Ushakov and G. Zapata at CRM, D. Grigoriev and D. Tieudjo at MPIM, and B. Burton, M. Elder and S. Tillmann in Australia. 
For further discussions during conferences I thank J. Gonzalez-Meneses, P. Bellingeri, V. Gebhardt, E. and S.J. Lee.
Particularly, I thank B. Tsaban for continuing discussions over the last years.

\section{Anshel-Anshel-Goldfeld key establishment} \label{AAGKE}
\subsection{AAG key establishment protocol for monoids} \label{AAGmon}

Here we use and describe a slightly generalized version of the {\it AAG key establishment protocol for monoids} \cite{AAG99}.  
Though it is easy to introduce further generalisations, the following notion will suffice for our purposes. \\
For this {\it general AAG key establishment protocol for monoids} we need sets $S_1, S_2$, two feasible monoids $(M,\cdot _M),(N,\cdot _N)$, and functions
\[ \beta _i: S_i\times M \longrightarrow N, \quad \gamma _i: S_i\times N \longrightarrow N, \quad \pi _i: S_i \longrightarrow M \quad (i=1,2) \]
which satisfy the following conditions: 

\begin{description}
\item[{\rm (1)}] For $i=1, 2$, $\beta _i(x,\cdot ):M\rightarrow N$ is for all $x\in S_i$ a monoid homomorphism, i.e. 
\[ \forall x\in S_i, y_1,y_2\in M: \quad \beta _i(x,y_1\cdot _M y_2)=\beta _i(x,y_1)\cdot _N\beta _i(x,y_2). \]
\item[{\rm (2)}] For $i=1, 2$, it is, in general, not feasible to determine a secret $x\in S_i$ from the knowledge of
$y_1,y_2,\ldots ,y_k\in M$ and $\beta _i(x,y_1), \ldots ,\beta _i(x,y_k)\in N$. 
\item[{\rm (3)}] For all $x \in S_1$ and $y\in S_2$: \quad $\gamma _1(x,\beta _2(y,\pi _1(x)))=\gamma _2(y,\beta _1(x,\pi _2(y)))$.
\end{description}

Furthermore Alice and Bob select elements $s_1,\ldots ,s_m$, $t_1,\ldots ,t_n\in M$. These elements are public, and
they define submonoids $S_A=\langle s_1,\ldots ,s_m\rangle $ and $S_B=\langle t_1,\ldots ,t_n\rangle $ of $M$.
Now Alice and Bob have to perform the following protocol steps:
\begin{enumerate}
\item Alice generates an element $a\in S_1$ such that $\pi _1(a)\in S_A$, and Bob chooses a $b\in S_2$ s.t. $\pi _2(b)\in S_B$.
\item Alice computes the elements $\beta _1(a,t_1), \ldots ,\beta _1(a,t_n)$ and publicly announces this list. This list is her public key.
Analogously Bob computes the elements $\beta _2(b,s_1), \ldots ,\beta _2(b,s_m)$ and publishes this list. 
\item Knowing that $\pi _1(a)=r_1\cdots r_k$ with $r_i\in \{s_1,\ldots ,s_m\}$ for some $k\in \mathbb{N}$ and $i=1,\ldots ,k$, Alice computes from Bob's public key $\beta _2(b,\pi _1(a))=\beta _2(b,r_1\cdots r_k) \stackrel{(1)}{=} \beta _2(b,r_1)\cdots \beta _2(b,r_k)$. \\
And Bob, knowing $\pi _2(b)=u_1\cdots u_{k'}$ with $u_j\in \{t_1,\ldots ,t_n\}$ for some $k'\in \mathbb{N}$ and $j=1,\ldots ,k'$, computes from Alice's public key $\beta _1(a,\pi _2(b))=\beta _1(a,u_1\cdots u_{k'}) \stackrel{(1)}{=} \beta _1(a,r_1)\cdots \beta _1(a,u_{k'})$. 
\item Alice computes $K_A=\gamma _1(a,\beta _2(b,\pi _1(a)))$, and symmetrically Bob computes $K_B=\gamma _2(b,\beta _1(a,\pi _2(b)))$.
\end{enumerate} 

Because of (3), the equivalence $K_A=K_B$ holds in the monoid $N$. Now any key extractor $\phi $ defined on the monoid $N$ provides a shared key $\phi (K_A)$.
Here a {\it key extractor} is any effectively computable function from a monoid to any keyspace\footnote{A standard key space is the semigroup of bitstrings $\{0, 1\}^*$.} (compare with \cite{AAG03}). A key extractor may be given by a normal form algorithm in the monoid, but in general the key extractor map needs not be injective. Anyway, for brevity we will refer in the sequel to the monoid element $K:=K_A \in N$ as the shared key. \\ 
Alice's secret key is the pair $(a,I)\in S_1\times \{1,\ldots ,m\}^k$ where $I$ denotes the index vector $(I_1,\ldots ,I_k)$ such that $r_i=s_{I_i}$ for $i=1,\ldots ,k$, i.e., $I$ determines a word over $\{s_1,\ldots ,s_m\}$ representing $\pi _1(a)\in S_A$. 
Analogously Bob's secret key is a pair $(b,J)\in S_2\times \{1,\ldots ,n\}^{k'}$ \\
The AAG key agreement scheme is formulated in a too general manner to be applied. For practical purposes we have to specify the sets $S_1, S_2$, the monoids $M,N$ and the functions $\beta _i,\gamma _i,\pi _i$ for $i=1, 2$. \\
Setting $S_1=S_2=M$, $\beta _1=\beta _2$ and $\pi _1=\pi _2={\rm id}_M$, we recover the original {\it AAG key establishment protocol for monoids} \cite{AAG99} as a special case of this generalized notion.

\subsection{AAG commutator KEP for groups} \label{AAGgr}
The {\it AAG commutator KEP for groups} \cite{AAG99} is determined by the following specifications: 
Let $S_1=S_2=M=N=G$ be a group, and $S_A$ and $S_B$ are assumed to be subgroups of $G$\footnote{
Now $r_i$ and $u_j$ are elements from $\{s_1^{\pm 1},\ldots ,s_m^{\pm 1}\}$ and $\{t_1^{\pm 1},\ldots ,t_n^{\pm 1}\}$, respectively.
}. We have $\pi _1=\pi _2={\rm id}_G$ and $\beta _1=\beta _2=:\beta $. The functions $\beta , \gamma _1,\gamma _2: G^2 \rightarrow G$ are defined by
\[ \beta (x,y)=x^{-1}yx, \quad \gamma _1(x,y)=x^{-1}y, \quad \gamma _2(x,y)=y^{-1}x. \]
Note that the shared key is the commutator
\begin{eqnarray*}
 K_A&=&\gamma _1(a,\beta _2(b,\pi _1(a)))=\gamma _1(a,b^{-1}ab)=a^{-1}(b^{-1}ab)=[a,b] \\ 
    &=& (a^{-1}ba)^{-1}b=\gamma _2(b,a^{-1}ba)=\gamma _2(b,\beta _1(a,\pi _2(b)))=K_B.
\end{eqnarray*}
If the group elements are given by representative words (over some alphabet of generators) as usual in combinatorial group theory, then multiplication is defined by simple concatenation of words. Therefore Alice and Bob have to publish the words representing the elements $\beta (a,t_i)=a^{-1}t_ia$ and $\beta (b,s_j)=b^{-1}s_jb$ in a disguised form. Therefore the question, whether one can efficiently disguise elements by using defining relations \cite{SZ06}, is very important for any platform group. One way is to use efficiently computable normal forms. \\
Such efficiently computable normal forms exist in many groups, e.g., in braid groups. Furthermore, the conjugator search, i.e. determining $x$ from $\beta (x,y)=x^{-1}yx$, was assumed to be hard in braid groups. Therefore Anshel, Anshel and Goldfeld suggested braid groups as platform groups for the AAG commutator KEP \cite{AAG99}. 

\subsection{Group Diffie-Hellman key establishment} \label{grDHKE}

In 2000 Ko, Lee, Cheo, Han, Kang and Park introduced a new key agreement scheme based on braid groups \cite{KL+00}. Here we describe a generalized version
of this KEP \cite{CK+01} for a general platform group $G$. Since this KEP is a non-abelian generalization of the classical Diffie-Hellman (DH) key agreement in the
abelian group $\mathbb{F}^{\times }_p$ \cite{DH76}, we call it the {\it group Diffie-Hellman (DH) key establishment protocol}. 
Let $(A_1, B_1)$ and $(A_2, B_2)$ be two pairs of public, commuting subgroups of a given group $G$, i.e.,  we have $[A_i, B_i]=1$ for $i=1, 2$. 
Furthermore, let $x$ be a ``generic'' element in $G$. Alice and Bob have to perform the following protocol steps.
\begin{enumerate}
\item Alice generates her secret key $(a_1, a_2) \in A_1 \times A_2$.
And Bob selects his private key $(b_1, b_2) \in B_1 \times B_2$.
\item Alice computes $y_A=a_1xa_2$ and sends it to Bob.
And Bob computes $y_B=b_1xb_2$ and submits it to Alice.
\item Alice receives $y_B$ and computes $K_A:=a_1y_Ba_2$.
And Bob receives $y_A$ and computes the shared key
\[ K_B:=b_1y_Ab_2=b_1(a_1xa_2)b_2=a_1(b_1xb_2)a_2=a_1y_Ba_2=K. \]
\end{enumerate}

For $a_1=a_2^{-1}$ and $b_1=b_2^{-1}$ we obtain the original {\it Ko-Lee et al. protocol} \cite{KL+00}. 
In \cite{AAG03} it is shown that the Ko-Lee et al. protocol may be seen also as an instance of the Anshel-Anshel-Goldfeld KEP for monoids. 
The following proposition is a straightforward generalization of that claim from \cite{AAG03} using the same proof idea\footnote{It is also a corrected
reformulation of Proposition 5.1 in \cite{Ka07}.}. 
\begin{prop} \hspace{-0.15cm}{\bf .} \label{grDH}
The group Diffie-Hellman key establishment protocol is an instance of the general Anshel-Anshel-Goldfeld KEP for monoids. 
\end{prop}
{\sc Proof.} Here we set $S_1=A_1\times A_2$, $S_2=B_1\times B_2$, $M=G$ and $N=\{ g_1xg_2 \mid (g_1,g_2)\in G^2 \}$. On $N$ we define the following ``forgetful'' binary opperation:
\[ 1\cdot _Nu=u\cdot _N1=u \quad (\forall x\in N) \quad {\rm and} \quad u\cdot _Nv=u \quad (\forall u,v\in N ,u\ne 1, v\ne 1). \]
This turns $N$ into a monoid. We define the functions $\beta _1: (A_1\times A_2)\times G \rightarrow N$ and $\beta _2: (B_1\times B_2)\times G \rightarrow N$ by
\[ \beta _1((u_1,u_2),v)=\beta _2((u_1,u_2),v)=u_1xu_2. \]
Then condition (1) is satisfied obviously. Indeed, given the forgetful operation on $N$, any constant function $\beta (u): G \rightarrow N$ provides
a monoid homomorphism. Further, condition (2) holds, because it is assumed to be hard for the group $G$ to determine $a=(a_1,a_2)\in A_1\times A_2$ from 
$\beta ((a_1,a_2),b)=a_1xa_2$. The computational problem is a search version of the {\it Double Coset Problem} or {\it Decomposition Problem} (DCP) - see also
section \ref{BaseProblems}.
We define the functions $\gamma _1: (A_1\times A_2)\times N \rightarrow N$ and $\gamma _2: (B_1\times B_2)\times N \rightarrow N$ by
\[ \gamma _1((u_1,u_2),v)=\gamma _2((u_1,u_2),v)=u_1vu_2. \]
Then (3) is satisfied, because we have for all $a=(a_1,a_2)\in A_1\times A_2$, $b=(b_1,b_2)\in B_1\times B_2$ (recall $[A_i, B_i]=1$ for $i=1, 2$):
\begin{eqnarray*} \gamma _1(a,\beta _2(b,\pi _1(a)))&=&\gamma _1((a_1,a_2),b_1xb_2)=a_1(b_1xb_2)a_2= \\
b_1(a_1xa_2)b_2&=&\gamma _2((b_1,b_2),a_1xa_2)=\gamma _2(b,\beta _1(a,\pi _2(b))). \end{eqnarray*}
This proves that the conditions (1)-(3) are fulfilled. It remains to show that the protocol steps 1.-3. of the DH-KEP are specializations of the protocol 
steps 1.-4. of the general AAG-KEP. Set $S_A=S_B=\langle x\rangle$ and define, for $i=1, 2$, $\pi _i: S_i \longrightarrow M$ by $u \mapsto x$, i.e., $\pi _1, \pi _2$ are 
constant functions.
\begin{enumerate}
\item Alice generates an element $a=(a_1,a_2)\in S_1=A_1\times A_2$ such that $\pi _1(a)=x\in \langle x\rangle=S_A$, and Bob chooses a $b=(b_1,b_2)\in S_2=B_1\times B_2$ s.t. $\pi _2(b)=x\in \langle x\rangle=S_B$.
\item Alice computes the element $\beta _1(a,x)=a_1xa_2=y_A$ and publicly announces this element. This element is her public key.
Analogously Bob computes the element $\beta _2(b,x)=b_1xb_2=y_B$ and publishes this element. 
\item Knowing that $\pi _1(a)=x$, Alice computes from Bob's public key $\beta _2(b,\pi _1(a))=\beta _2(b,x)=b_1xb_2$. Indeed, this computation is trivial, because
here $\beta _2(b,\pi _1(a))$ is Bob's public key. \\
And Bob, knowing $\pi _2(b)=x$, computes from Alice's public key $\beta _1(a,\pi _2(b))=\beta _1(a,x)=a_1xa_2$. Also this computation is trivial. Therefore, here protocol step 3 becomes redundant.   
\item Alice computes $K_A=\gamma _1(a,\beta _2(b,\pi _1(a)))=a_1y_Ba_2$, and symmetrically Bob computes $K_B=\gamma _2(b,\beta _1(a,\pi _2(b)))=b_1y_Ab_2$. \quad
$\Box $
\end{enumerate}

We have proven that the group DH-KEP is a special case of the AAG-KEP for monoids. Nevertheless, not every special case is obvious. Indeed, the group DH-KEP does not use the homomorphy property (1) at all. Therefore step 3 in this specification of the general AAG-KEP (see proof above) became trivial. 
This observation motivates us to introduce the following somehow informal notion: 
\begin{defi}
{\rm We call a key establishment protocol {\it AAG-like} if it is an instance of
the general AAG-KEP {\it and} it utilizes property (1) in a non-trivial way.}  
\end{defi}
According to this notion, and contrary to the AAG commutator KEP, the group Diffie-Hellman KEP is {\it not} an AAG-like KEP, though it can be formally considered as an instance of the general AAG-KEP (see Proposition \ref{grDH}).

\subsection{Base Problems} \label{BaseProblems}

The following search problems are related with the group based protocols from the previous sections. Let $G$ be a group.
\begin{list}{}{\setlength{\itemsep}{0cm} \setlength{\parsep}{0cm} }
\item[{\bf CSP} ({\it Conjugacy Search Problem}):]
\item[{\sc Input:}] $(s,s^x) \in G^2$. ($s^x$ denotes $x^{-1}sx$.)
\item[{\sc Objective:}] Find $x' \in G$ such that $s^{x'}=s^x$. 
\\
\item[{\bf $l$-simCSP} ({\it $l$-Simultaneous Conjugacy Search Problem}):]
\item[{\sc Input:}] $\{ (s_i,s_i^x) \in G^2 | i=1,\ldots ,l\}$.
\item[{\sc Objective:}] Find $x' \in G$ such that $s_i^{x'}=s_i^x \quad \forall i=1,\ldots ,l$. 
\\
\item[{\bf subCSP} ({\it Subgroup Conjugacy Search Problem}):] Let $H$ be a subgroup of $G$.
\item[{\sc Input:}] $(s,s^x) \in G^2$ with $x \in H \le G$.
\item[{\sc Objective:}] Find $x' \in H$ such that $s^{x'}=s^x$. 
\\
\item[{\bf $l$-ssCSP} ({\it $l$-Simultaneous Subgroup Conjugacy Search Problem}):]
\item[{\sc Input:}] $\{ (s_i,s_i^x) \in G^2 | i=1,\ldots ,m \}$ with $x \in H \subset G$.
\item[{\sc Objective:}] Find $x' \in H$ such that $s_i^{x'}=s_i^x \quad \forall i=1,\ldots ,l$. 
\\
\item[{\bf AAGP} ({\it Anshel-Anshel-Goldfeld Problem}):] 
Let $A=\langle a_1,\ldots ,a_k\rangle $ and $B=\langle b_1, \ldots ,b_m\rangle $ be two f.g. subgroups of $G$.
\item[{\sc Input:}] $\{ (a_i,a_i^y) \in G^2 | i=1,\ldots ,k\} \cup \{ (b_j,b_j^x) \in G^2 | j=1,\ldots ,m\}$ with $x \in A$ 
and $y \in B$.
\item[{\sc Objective:}] Find $K:=x^{-1}y^{-1}xy$. 
\\
\item[{\bf KLP} ({\it Ko-Lee Problem - a Diffie-Hellman version of the GCSP or CDP}):] Let $A,B \le G$ with $[A,B]=1$.
\item[{\sc Input:}] $(s,s^x,s^y) \in G^3$ with $x \in A, y \in B$.
\item[{\sc Objective:}] Find $K:=x^{-1}y^{-1}sxy$. 
\\
\item[{\bf DCP} ({\it Double Coset or Decompositon Problem}):] Let $H_1, H_2 \le G$.
\item[{\sc Input:}] $(s,x_1sx_2) \in G^2$ for some $x_1 \in H_1$ and $x_2\in H_2$.
\item[{\sc Objective:}] Find $(x'_1,x'_2) \in H_1 \times H_2$ such that $x'_1sx'_2=x_1sx_2$. 
\\
\item[{\bf CDP} ({\it Conjugacy Decompositon Problem}):]
\item[{\sc Input:}] $(s,s^x) \in G^2$ with $x \in H \le G$.
\item[{\sc Objective:}] Find $(x'_1,x'_2) \in H^2$ such that $x'_1sx'_2=s^x$. 
\\
\item[{\bf DH-DCP} ({\it Diffie-Hellman Decompositon Problem}):] $A_1, A_2, B_1, B_2$ subgroups of $G$ such that $[A_i,B_i]=1$ for $i=1, 2$.
\item[{\sc Input:}] $(s,x_1sx_2,y_1sy_2) \in G^3$ with $x_1\in A_1$, $x_2 \in A_2$, $y_1\in B_1$, $y_2 \in B_2$.
\item[{\sc Objective:}] Find $K:=x_1y_1sx_2y_2$. 
\end{list}
Indeed, the AAG commutator KEP, the Ko-Lee protocol and the group DH-KEP are based on the AAGP, KLP and DH-DCP, respectively. \\
Now, let $P_1, P_2$ be two computational problems. We say $P_1$ is harder than $P_2$ or $P_1$ implies $P_2$, written $P_1\rightarrow P_2$, if a $P_1$-oracle
provides a solution to problem $P_2$. \\

\begin{prop} \hspace{-0.15cm}{\bf .} \label{BasePs}
We have the following hierarchy of search problems:
\end{prop}

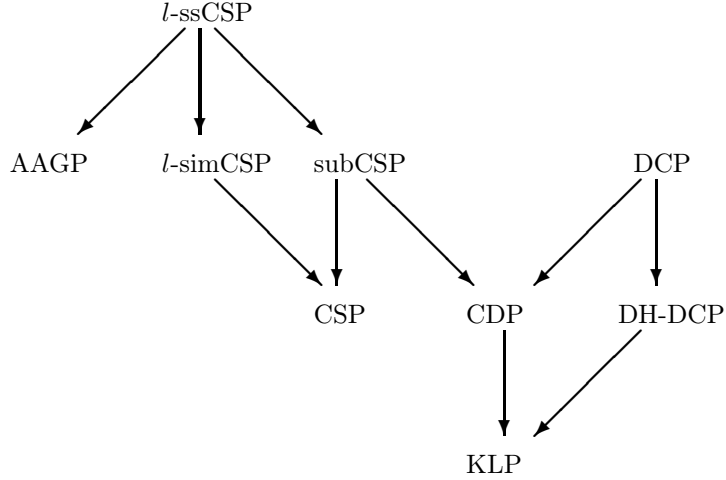
\setlength{\unitlength}{1cm}
\begin{picture}(10,7)

\put(2,6){$l$-ssCSP}
\put(0,4){AAGP}
\put(2,4){$l$-simCSP}
\put(4,4){subCSP}
\put(8.2,4){DCP}
\put(4,2){CSP}
\put(6,2){CDP}
\put(8,2){DH-DCP}
\put(6,0){KLP}

\thicklines

\put(2.3,5.9){\vector(-1,-1){1.4}}
\put(2.5,5.9){\vector(0,-1){1.4}}
\put(2.7,5.9){\vector(1,-1){1.4}}
\put(2.7,3.9){\vector(1,-1){1.4}}
\put(4.3,3.9){\vector(0,-1){1.4}}
\put(4.7,3.9){\vector(1,-1){1.4}}
\put(8.3,3.9){\vector(-1,-1){1.4}}
\put(8.5,3.9){\vector(0,-1){1.4}}
\put(6.5,1.9){\vector(0,-1){1.4}}
\put(8.3,1.9){\vector(-1,-1){1.4}}

\end{picture}

{\sc Proof.} Most of the sketched implications are obvious consequences of the definitions. We just prove CDP $\rightarrow $ KLP and $l$-ssCSP $\rightarrow $ AAGP:
\begin{enumerate}
\item (see \cite{KL+00}) The input is a triple $(s,s^x,s^y) \in G^3$ with $x \in A, y \in B$, and $A,B \subset G$ with $[A,B]=1$.
A CDP-oracle provides $(x_1,x_2)\in A^2$ with $x_1sx_2=s^x$. Now we can compute the shared key
\[ x_1s^yx_2=x_1y^{-1}syx_2=y^{-1}(x_1sx_2)y=y^{-1}(x^{-1}sx)y=K. \]
\item Here the input is $\{ (a_i,a_i^y) \in G^2 | i=1,\ldots ,k\} \cup \{ (b_j,b_j^x) \in G^2 | j=1,\ldots ,m\}$ with $x \in A=\langle a_1,\ldots ,a_k\rangle $ 
and $y \in B=\langle b_1, \ldots ,b_m\rangle $.
A $m$-SGCSP-oracle provides a $x'\in A$ with $x'^{-1}b_jx'=b_j^{x}$ for all $j=1,\ldots ,m$. 
And a $k$-SGCSP-oracle provides a $y'\in B$ with $y'^{-1}a_iy'=a_i^{y}$ for all $i=1,\ldots ,k$.
Now, since $x'^{-1}b_jx'=b_j^{x} \Leftrightarrow [x'x^{-1}, b_j]=1 \, \forall j$, we have
$x'=c_bx$ for some $c_b\in C_G(B)$. Here $C_G(B)$ denotes the intersection of all centralizers $C_G(b_j)$ of $b_j$ ($j=1,\ldots ,m$) in $G$. 
Analogously, we can write $y'=c_ay$ with $c_a\in C_G(A)=\bigcap _{i=1}^k C_G(a_i)$. \\
Now, $x'\in A$ implies $c_b\in A$. Therefore we have $[c_a,c_b]=1$, and we can compute the shared key
\begin{eqnarray*}
K':=x'^{-1}y'^{-1}x'y'=(c_bx)^{-1}(c_ay)^{-1}c_bxc_ay&=&x^{-1}c_b^{-1}y^{-1}c_a^{-1}c_bxc_ay \\
           =x^{-1}y^{-1}c_b^{-1}c_a^{-1}c_bc_axy &\stackrel{!}{=}& x^{-1}y^{-1}xy=K.  \quad \Box
\end{eqnarray*}
\end{enumerate}

We see, that solving the classical CSP is insufficient for breaking the AAG protocol or the Ko-Lee protocol. Furthermore, it is, in general, insufficient to solve 
the $l$-SCSP to obtain the shared key $K$ of the AAG protocol \cite{SU06}: \\
Let $x'=c_bx\in G$ and $y'=c_ay\in G$ with $c_a\in C_G(A), c_b\in C_G(B)$ be the output of a $m$-SCSP-oracle and a $k$-SCSP-oracle, respectively.
Then we have $K'=K$ if and only if $[c_a,c_b]=1$. A necessary condition for $[c_b,c_a] \ne 1$ is $c_b \notin A \wedge c_a \notin B$, which implies 
$x' \notin A \wedge y' \notin B$. Otherwise, if $x' \notin A$, but $y' \in B$ (or vice versa), the adversary gets $K'=K$. \\
Alternatively, the adversary could solve the SCSP and the
\begin{list}{}{\setlength{\itemsep}{0cm} \setlength{\parsep}{0cm} }
\item[{\bf MSP} ({\it Membership Search Problem}):]
\item[{\sc Input:}] $x, a_1,\ldots ,a_k\in G$.
\item[{\sc Objective:}] Find an expression of $x$ as a word in $a_1,\ldots ,a_k$ (notation $x=x(a_1,\ldots ,a_k)$), if it exists, i.e. if $x\in \langle a_1,\ldots ,a_k \rangle $.
\end{list}
to break the AAG key agreement scheme \cite{SU06}: \\
If a $m$-SCSP-oracle outputs a $x'=c_bx\in A$, then the MSP-oracle provides the word expression $x'(a_1,\ldots ,a_k)$. Now the adversary can compute
the shared key
\[ x'^{-1}x'(a_1^y,\ldots ,a_k^y)=x'^{-1}x'^y=(x^{-1}c_b)y^{-1}(c_bx)y=[x,y]=K. \]
But we have shown above, that it is not necessary to solve the MSP.

\subsection{Diffie-Hellman key establishment protocol} \label{DHKE}

Recall the classical {\it Diffie-Hellman key establishment protocol} \cite{DH76}. Let $G$ be a cyclic group and $x$ an element of big order in $G$.
Alice and Bob have to perform the following protocol steps.
\begin{description}
\item[{\rm 1.}]  Alice chooses a $k\in \mathbb{Z}$, computes $y_A=x^k$, and sends it to Bob.
And Bob chooses a $l\in \mathbb{Z}$, computes $y_B=x^l$, and submits it to Alice.
\item[{\rm 2.}] Alice receives $y_B$ and computes $K_A:=y_B^k$.
And Bob receives $y_A$ and computes the shared key $K_B:=y_A^l=(x^k)^l=(x^l)^k=y_B^k=K_A$.
\end{description}

\begin{prop} \hspace{-0.15cm}{\bf .} \label{PropDH}
The Diffie-Hellman key establishment protocol is an instance of the Anshel-Anshel-Goldfeld KEP for monoids. Furthermore it is a AAG-like KEP.
\end{prop}
{\sc Proof.} Here we set $S_1=S_2=\mathbb{Z}$ and $M=N=S_A=S_B=\langle x\rangle$. For $i=1, 2$, we define the functions $\beta _i, \gamma _i: \mathbb{Z}\times \langle x\rangle \rightarrow \langle x\rangle$ and $\pi _i: \mathbb{Z} \rightarrow \langle x\rangle$ by 
\[ \beta _i(k,y)=y^k, \quad \gamma _i(k,y)=y \quad {\rm and} \quad \pi _i(k)=x^k. \]
Then, for $i=1, 2$, condition (1) holds for all $y_1, y_2\in M$, because $M=\langle x\rangle$ is cyclic, and therefore abelian:
\[ \beta _i(k,y_1\cdot y_2)=(y_1y_2)^k=y_1^ky_2^k=\beta _i(k,y_1)\cdot \beta _i(k,y_2). \] 
Note that exponentiation is only a homomorphism if the monoid $M$ is abelian.
Further, condition (2) holds, because it is assumed to be hard to determine $k \in \mathbb{Z}$ from 
$\beta (k,x)=x^k$. The computational problem is well known as the {\it Discrete Logarithm Problem} (DLP). \\
And (3) is satisfied, because we have for all $k, l\in \mathbb{Z}$:
\[ \gamma _1(k,\beta _2(l,\pi _1(k)))=\beta _2(l,x^k)=(x^k)^l=(x^l)^k=\beta _1(k,x^l)=\gamma _2(l,\beta _1(k,\pi _2(l))). \]
This proves that the conditions (1)-(3) are fulfilled. It remains to show that the protocol steps 1.-2. of the Diffie-Hellman KEP are specializations of the protocol steps 1.-4. of the general AAG-KEP. 
\begin{enumerate}
\item Alice generates an element $k\in S_1=\mathbb{Z}$ such that $\pi _1(k)=x^k\in \langle x\rangle=S_A$, and Bob chooses a $l\in S_2=\mathbb{Z}$ s.t. $\pi _2(l)=x^l\in \langle x\rangle=S_B$.
\item Alice computes the element $\beta _1(k,x)=x^k=y_A$ and publicly announces this element. This element is her public key.
Analogously Bob computes the element $\beta _2(l,x)=x^l=y_B$ and publishes this element. 
\item Knowing that $\pi _1(k)=x^k$, Alice computes from Bob's public key $\beta _2(l,\pi _1(k))=\beta _2(l,x^k)=(x^k)^l=(x^l)^k=y_B^k$. 
And Bob, knowing $\pi _2(l)=x$, computes from Alice's public key $\beta _1(k,\pi _2(l))=\beta _1(k,x^l)=(x^l)^k=(x^k)^l=y_A^l$. 
\item Alice computes $K_A=\gamma _1(k,\beta _2(l,\pi _1(k)))=\beta _2(l,\pi _1(k))=y_B^k$, and symmetrically Bob computes $K_B=\gamma _2(l,\beta _1(k,\pi _2(l)))=\beta _1(k,\pi _2(l))=y_A^l$. Since this is exactly the output of the computation in step 3, here step 4 is redundant or trivial.
\end{enumerate}
Let us recall and emphasize that in step 3 the homomorphy property (1) is used in a nontrivial way. For example,  Alice knowing
$\pi _1(k)=x^k=\underbrace{x\cdots x}_{k \,\,{\rm times}}=:w_k(x)$ can compute
\begin{eqnarray*}
y_B^k&=&(\beta _2(l,x))^k=(x^l)^k=w_k(x^l)=\underbrace{x^l\cdots x^l}_{k \,\,{\rm times}}\stackrel{(1)}{=}(\underbrace{x\cdots x}_{k \,\,{\rm times}})^l \\
&=&(w_k(x))^l=(x^k)^l=\beta _2(l,w_k(x))=\beta _2(l,\pi _1(k)).
\end{eqnarray*}
Therefore, we may view the classical DH-KEP as an AAG-like KEP. \quad $\Box $ 

\subsection{Sakalauskas, Tvarijonas and Raulynaitis Key Establishment Protocol (STR-KEP)} \label{STRKEP}

The following KEP is a natural hybrid of the classical DH-KEP and the Ko-Lee-KEP. It was introduced in 2007 by Sakalauskas, Tvarijonas and Raulynaitis in \cite{STR07}  \\
Let $G$ be a (noncommutative) group and $A, B$ a pair of commuting subgroups in $G$. 
Furthermore, let $x$ be a ``generic'' element in $G$. Alice and Bob have to perform the following protocol steps.
\begin{enumerate}
\item Alice generates her secret key $(k, a) \in \mathbb{Z} \times A$.
And Bob selects his private key $(l, b) \in \mathbb{Z} \times B$.
\item Alice computes $y_A=a^{-1}x^ka$ and sends it to Bob.
And Bob computes $y_B=b^{-1}x^lb$ and submits it to Alice.
\item Alice receives $y_B$ and computes $K_A:=a^{-1}y_B^ka$.
And Bob receives $y_A$ and computes the shared key
\begin{eqnarray*}
K_B&:=&b^{-1}y_A^lb=b^{-1}(a^{-1}x^ka)^lb=b^{-1}(a^{-1}(x^k)^la)b \\
&=&a^{-1}(b^{-1}(x^l)^kb)a=a^{-1}(b^{-1}x^lb)^ka=a^{-1}y_B^ka=K_A. 
\end{eqnarray*} 
\end{enumerate}

\begin{prop} \hspace{-0.15cm}{\bf .} \label{PropSTRKEP}
The Sakalauskas, Tvarijonas and Raulynaitis Key Establishment Protocol is an instance of the Anshel-Anshel-Goldfeld KEP for monoids. Furthermore, it is an AAG-like KEP.
\end{prop}
{\sc Proof.} Here we set $S_1=\mathbb{Z} \times A$, $S_2=\mathbb{Z} \times B$, $M=S_A=S_B=\langle x\rangle$ and $N=G$. For $i=1, 2$, we define the functions $\beta _i: S_i\times \langle x\rangle \rightarrow G$ and $\pi _i: S_i \rightarrow \langle x\rangle$ by 
\[ \beta _i((k,z),y)=z^{-1}y^kz, \quad \gamma _i((k,z),y)=z^{-1}yz \quad {\rm and} \quad \pi _i(k)=x^k. \]
Then, for $i=1, 2$, condition (1) holds for all $y, y'\in M$:
\[ \beta _i((k,z),y\cdot y')=z^{-1}(yy')^kz=z^{-1}y^kz \cdot z^{-1}y'^kz=\beta _i((k,z),y)\cdot \beta _i((k,z),y'). \] 
Further, condition (2) holds, because it is assumed to be hard to determine $k \in \mathbb{Z}$ and $z\in G$ from 
$\beta ((k,z),x)=z^{-1}x^kz$. The computational problem is a ``mixed problem'' requiring to solve simultaneously the DLP and the CSP (see \cite{STR07}). \\
And (3) is satisfied, because we have for all $k, l\in \mathbb{Z}$, $a\in A$, $b\in B$:
\begin{eqnarray*} \gamma _1((k,a),\beta _2((l,b),\pi _1(k,a)))&=&a^{-1}\beta _2((l,b),x^k)a=a^{-1}(b^{-1}(x^k)^lb)a \\
=b^{-1}a^{-1}(x^l)^kab &=& b^{-1}\beta _1((k,a),x^l)b=\gamma _2((l,b),\beta _1((k,a),\pi _2(l,b))). 
\end{eqnarray*}
This proves that the conditions (1)-(3) are fulfilled. It remains to show that the protocol steps 1.-2. of the Diffie-Hellman KEP are specializations of the protocol steps 1.-4. of the general AAG-KEP. 
\begin{enumerate}
\item Alice generates an element $(k,a)\in \mathbb{Z}\times A$ such that $\pi _1(k,a)=x^k\in \langle x\rangle=S_A$, and Bob chooses $(l,b)\in \mathbb{Z} \times B$ s.t. $\pi _2(l,b)=x^l\in \langle x\rangle=S_B$.
\item Alice computes the element $\beta _1((k,a),x)=a^{-1}x^ka=y_A$ and publicly announces this element. This element is her public key.
Analogously Bob computes the element $\beta _2((l,b),x)=b^{-1}x^lb=y_B$ and publishes this element. 
\item Knowing that $\pi _1(k,a)=x^k$, Alice computes from Bob's public key $\beta _2((l,b),\pi _1(k,a))=\beta _2((l,b),x^k)=b^{-1}(x^k)^lb=(b^{-1}x^lb)^k=y_B^k$. 
And Bob, knowing $\pi _2(l,b)=x$, computes from Alice's public key $\beta _1((k,a),\pi _2(l,b))=\beta _1((k,a),x^l)=a^{-1}(x^l)^ka=(a^{-1}x^ka)^l=y_A^l$. 
\item Alice computes $K_A=\gamma _1((k,a),\beta _2((l,b),\pi _1(k,a)))=a^{-1}\beta _2((l,b),\pi _1(k,a))a=a^{-1}y_B^ka$, and symmetrically Bob computes $K_B=\gamma _2((l,b),\beta _1((k,a),\pi _2(l,b)))=
b^{-1}\beta _1((k,a),\pi _2(l,b))b=a^{-1}y_A^lb$. 
\end{enumerate}
Let us recall and emphasize that also here in step 3 the homomorphy property (1) is used in a nontrivial way. For example,  Alice knowing
$\pi _1(k,a)=x^k=\underbrace{x\cdots x}_{k \,\,{\rm times}}=:w_k(x)$ can compute
\begin{eqnarray*}
y_B^k&=&(\beta _2((l,b),x))^k=(b^{-1}x^lb)^k=w_k(b^{-1}x^lb)=\underbrace{b^{-1}x^lb\cdots b^{-1}x^lb}_{k \,\,{\rm times}} \\
& \stackrel{(1)}{=} & b^{-1}(\underbrace{x\cdots x}_{k \,\,{\rm times}})^lb=b^{-1}(w_k(x))^lb=b^{-1}(x^k)^lb=\beta _2((l,b),\pi _1(k,a)).
\end{eqnarray*}
Therefore, we may view the STR-KEP as an AAG-like KEP. \quad $\Box $

\section{Key establishment using non-associative operations}  \label{nassKE}
\subsection{AAG scheme for magmas} \label{AAGmagma}  

Monoids are proposed as algebraic platform structures for the AAG key agreement protocol in \cite{AAG99}. But the monoid structure is only used in the AAG scheme in order to guarantee that the secret key, e.g. Alice's key $a$, is an uniquely defined product of some given generators $\{s_1,\ldots ,s_m\}$, i.e. $a=r_1\cdot r_2\cdots r_k$
with $r_i\in \{s_1,\ldots ,s_m\}$ for all $i$. It is, of course, no problem to introduce brackets in this expression in order to handle nonassoziative operations. Therefore, there exists a straightforward generalization of the AAG scheme from monoids to magmas. \\
A {\it magma} (sometimes also called grupoid) $(M,*)$ is a set $M$ equipped with a binary operation $*$ on $M$, i.e. a function $M\times M\rightarrow M$. 
Note that there are no relations, which have to be satisfied by the elements of $M$. The notion of a magma was introduced by N. Bourbaki (see, e.g., \cite{Bo74}). \\
We describe the AAG key establishment protocol in the - for our purposes - most general manner. \\
For $i=1,2$, let $S_i$ be a sets and $(M,\bullet _i)$ and $(N,\circ _i)$ be magmas, i.e. there are two operations on the sets $M,N$, respectively.
For $i=1,2$, we need functions 
\[ \beta _i: S_i\times M \rightarrow N, \quad \gamma _i: S_i\times N\rightarrow N, \quad \pi _i: S_i\rightarrow M \]
which satisfy the following three conditions:
io
\begin{description}
\item[{\rm (1)}] $\beta _1 (x,\cdot ): (M, \bullet _2) \rightarrow (N, \circ _2)$ is for all $x\in S_1$ a {\it magma homomorphism}\footnote{
More on magmas and magma homomorphisms can be found, e.g. in \cite{Se65,Ge94}.}, i.e. 
\[ \forall x\in S_1,y, y'\in M: \quad \beta _1(x,y\bullet _2 y')=\beta _1(x,y)\circ _2\beta _1(x,y'). \]
Also $\beta _2 (x,\cdot ): (M, \bullet _1) \rightarrow (N, \circ _1)$ is for all $x\in S_2$ a magma morphism, i.e. 
\[ \forall x\in S_2,y, y'\in M: \quad \beta _2(x,y\bullet _1 y')=\beta _2(x,y)\circ _1\beta _2(x,y'). \]
\item[{\rm (2)}] It is, in general, not feasible to determine a secret $x\in S_i$ ($i=1, 2$) from the knowledge of
\[ y_1, y_2 ,\ldots ,y_k\in M \quad {\rm and} \quad \beta _i(x,y_1),\beta _i(x,y_2), \ldots ,\beta _i(x,y_k). \]
\item[{\rm (3)}] For all $a \in S_1, b\in S_2: \quad \gamma _1(a,\beta _2(b,\pi_1(a)))=\gamma _2(b,\beta _1(a,\pi_2(b)))$.
\end{description}

Consider an element $y$ of a magma $(M,\bullet )$ which is an iterated product of other elements in $M$. Such an element can be described by a planar rooted binary
tree $T$ whose $k$ leaves are labelled by these other elements $y_1,\ldots ,y_k\in M$. We use the notation $y=T_{\bullet }(y_1,\ldots ,y_k)$. 
Here the subscript $\bullet $ tells us that the grafting of subtrees of $T$ corresponds to the operation $\bullet $. \\
Now, it is easy to prove by induction that any magma homomorphism $\beta :(M,\bullet )\rightarrow (N,\circ )$ satisfies
\[ \beta (T_{\bullet }(y_1,\ldots ,y_k))=T_{\circ }(\beta (y_1),\ldots ,\beta (y_k)) \]
for all $y_1,\ldots ,y_k\in M$. In particular, the magma morphisms $\beta _1(x,\cdot ),\beta _2(x,\cdot )$ ($x\in S$) fulfill this property. \\
Alice and Bob publicly assign sets $\{s_1,\ldots ,s_m\}, \{t_1,\ldots ,t_n\}\subset M$, respectively. The secret key spaces $SK_A, SK_B$ of Alice and Bob are subsets of $S_1, S_2$,
respectively, and they depend on these public elements. 
It is sufficient that $\beta _1,\beta _2$ fulfill condition $(1)$ only for all $x\in SK_A,SK_B$, respectively, and that condition $(3)$ holds for all $a\in SK_A,b\in SK_B$. \\
Now, Alice and Bob perform the following protocol steps.

\begin{description}
\item[{\rm 1.}] Alice generates her secret key $a\in SK_A$, and Bob chooses his secret key $b\in SK_B$.
\item[{\rm 2.}] Alice computes the elements $\beta _1(a,t_1), \ldots ,\beta _1(a,t_n)\in N$, and sends them to Bob.
Analogously Bob computes the elements $\beta _2(b,s_1), \ldots ,\beta _2(b,s_m)\in N$, and sends them to Alice. 
\item[{\rm 3.}] Alice, knowing $\pi _1(a)=T_{\bullet _1}(r_1, \ldots , r_k)$ with $r_i\in \{s_1,\ldots ,s_m\}$, computes from Bob's public key
\[ T_{\circ _1}(\beta _2(b,r_1), \ldots , \beta _2(b,r_k))=\beta _2(b,T_{\bullet _1}(r_1, \ldots , r_k))=\beta _2(b,\pi _1(a)). \]
And Bob, knowing $p_2(b)=T'_{\bullet _2}(u_1, \ldots , u_{k'})$ with $u_j\in \{t_1,\ldots ,t_n\}$, computes from Alice's public key
\[ T'_{\circ _2}(\beta _1(a,u_1), \ldots , \beta _1(a,u_{k'}))=\beta _1(a,T'_{\bullet _2}(u_1, \ldots , u_{k'}))=\beta _1(a,\pi _2(b)). \]
\item[{\rm 4.}] Alice computes $K:=\gamma _1(a,\beta _2(b,\pi _1(a)))$.
Bob also computes the shared key $\gamma _2(b,\beta _1(a,\pi _2(b))) \stackrel{(3)}{=}K$.
\end{description} 

Note that the protocols described in section \ref{AAGmon} are special instances of this general AAG like protocol for magmas. \\
A natural special case of this scheme is given by $M=N=S_1=S_2$. This implies that the functions $\beta _i, \gamma _i$, for $i=1,2$,
induce further binary operations on $M$. If additionally $\bullet _i=\circ _i$ holds for $i=1,2$, then $M$ satisfies some distributive laws.
This will lead to the notion of LD- and multi-LD-systems (see section \ref{LD}). \\
Another specification of our general magma-based scheme is discussed in the next subsection.

\subsection{Non-associative KEP based on simultaneous DCP}  \label{simDCP}
\subsubsection{Specifications}

We consider the following specifications of the AAG scheme for magmas: \\
Let $G=M=N$ be a group, and set $S_1=S_2=G^2$. The group multiplication symbol in $G$ will usually be omitted.  
The operations $\bullet _i,\circ _i$ ($i=1,2$) on $G$ are defined by
\[ x\bullet _1y=x\bullet _2y=x\circ _1y=x\circ _2y\equiv x\bullet y:=xy^{-1}x, \]
and the functions $\beta _1,\beta _2: G^2\times G\rightarrow G$ are defined by
\[ \beta _1((x_1,x_2),y)=\beta _2((x_1,x_2),y)\equiv \beta ((x_1,x_2),y):=x_1yx_2. \]
$\beta (x,\cdot )$ fulfills the homomorphy condition $(1)$, for all $x=(x_1,x_2)\in G^2$, because
\begin{eqnarray*} \beta ((x_1,x_2),y_1)\bullet \beta ((x_1,x_2),y_2) &=& (x_1y_1x_2)\bullet (x_1y_2x_2)= \\
(x_1y_1x_2)x_2^{-1}y_2^{-1}x_1^{-1}(x_1y_1x_2) &=& x_1(y_1y_2^{-1}y_1)x_2=\beta ((x_1,x_2),y_1\bullet y_2). 
\end{eqnarray*} 
Alice and Bob publicly assign sets $\{s_1,\ldots ,s_m\}, \{t_1,\ldots ,t_n\}\subset G$, respectively. The secret key spaces of Alice and Bob are $SK_A=G\times S_A$ and
$SK_B=S_B\times G$, where $S_A=\langle s_1,\ldots ,s_m\rangle _{\bullet }$ and $S_B=\langle t_1,\ldots ,t_n\rangle _{\bullet }$ denote submagmas of 
$(G,\bullet )$ generated by the publicly assigned elements. \\
The projections $\pi_1, \pi_2:G^2\rightarrow G$ and the functions $\gamma _1,\gamma _2:G^2\times G\rightarrow G$ are defined by 
\[ \pi_1(x,y)=y, \,\, \pi_2(x,y)=x \quad {\rm and} \quad \gamma _1((x_1,x_2),y)=x_1y, \,\, \gamma _2((x_1,x_2),y)=yx_2. \]
These definitions satisfy condition $(3)$, because
\begin{eqnarray*} && \gamma _1(a, \beta (b,\pi_1(a)))=\gamma _1(a,\beta (b,a_r))=\gamma _1(a,b_la_rb_r)=a_l(b_la_rb_r) \\
&=& (a_lb_la_r)b_r=\gamma _2(b,a_lb_la_r)=\gamma _2(b,\beta (a,b_l))=\gamma _2(b,\beta (a,\pi_2(b)))
\end{eqnarray*}
for all $a=(a_l,a_r),b=(b_l,b_r)\in G^2$. \\
We skip repeating all the protocol steps from section \ref{AAGmagma} with these specifications. The base problem for these non-associative scheme is discussed in the next subsubsection.

\subsubsection{A related non-commutative scheme} \label{relNCsch}
Consider the right part of Alice's key $a_r=T_{\bullet }(r_1,\ldots ,r_k)\in S_A$ with $r_i\in \{s_1,\ldots ,s_m\}$. 
If we view $a_r$ as a word in the $s_i$'s, then we observe that $a_r$ is self-reverse and the exponent signs of $a_r$ alternate, beginning and ending with a positive sign.
For example, we have
\[ (r_1\bullet r_2)\bullet (r_3\bullet (r_4\bullet r_5))=r_1r_2^{-1}r_1r_3^{-1}r_4r_5^{-1}r_4r_3^{-1}r_1r_2^{-1}r_1. \]
While in this scheme alternating exponent signs are essential to gurantee that condition $(1)$ holds, the self-reverse property seems to be superflous.
It comes from the self-reverse property of the non-associative operation $\bullet $. 
Anyway, for example in order to compute $b_la_rb_r$, Alice actually doesn't need to know $a_r$ as a tree-word in the submagma $\langle s_1, \ldots ,s|m \rangle _{\bullet }$.
Rather it suffices to know $a_r$ as an ``alternating'' word of the form $s_{i_1}s_{i_2}^{-1}s_{i_3} \cdots s_{i_{2l}}^{-1}s_{i_{2l+1}}$.  \\
Therefore, we give up this restricted key choice and define modified (bigger) secret key spaces by $SK_A=G\times SK_A^{(r)}$ and $SK_B=SK_B^{(l)} \times G$ with
\begin{eqnarray*} SK_A^{(r)}&=&\{ r_1r_2^{-1}r_3r_4^{-1}\cdots r_{2l}^{-1}r_{2l+1} \mid r_i\in \{s_1,\ldots ,s_m\} \, \forall 1\le i\le l, l\in \mathbb{N}\}, \\
SK_B^{(l)}&=&\{ u_1u_2^{-1}u_3u_4^{-1}\cdots u_{2l'}^{-1}u_{2l'+1} \mid u_j\in \{t_1,\ldots ,t_n\} \, \forall 1\le j\le l', l'\in \mathbb{N}\}.
\end{eqnarray*}
Then, Alice and Bob have to perform the following protocol steps.
\begin{description}
\item[{\rm 1.}]  Alice generates her secret key $(a_l,a_r)\in G\times SK_A^{(r)}$. Bob chooses his secret key $(b_l,b_r)\in SK_B^{(l)}\times G$.
\item[{\rm 2.}]  Alice computes the elements $a_lt_1a_r, \ldots ,a_lt_na_r$, and sends them to Bob. Analogously Bob computes the elements $b_ls_1b_r, \ldots ,b_ls_mb_r$, and sends them to Alice. 
\item[{\rm 3.}]  Alice, knowing $a_r=r_1r_2^{-1}r_3r_4^{-1}\cdots r_{2l}^{-1}r_{2l+1}$ with $r_i\in \{s_1,\ldots ,s_m\}$, computes from Bob's public key
\begin{eqnarray*} && (b_lr_1b_r)(b_lr_2b_r)^{-1}(b_lr_3b_r)\cdots (b_lr_{2l}b_r)^{-1}(b_lr_{2l+1}b_r) \\
&=& b_l(r_1r_2^{-1}r_3\cdots r_{2l}^{-1}r_{2l+1})b_r=b_la_rb_r. \end{eqnarray*}
Bob, knowing $b_l=u_1u_2^{-1}u_3u_4^{-1}\cdots u_{2l'}^{-1}u_{2l'+1}$ with $u_j\in \{t_1,\ldots ,t_n\}$, computes from Alice's public key
\begin{eqnarray*} && (a_lu_1a_r)(a_lu_2a_r)^{-1}(a_lu_3a_r)\cdots (a_lu_{2l'}a_r)^{-1}(a_lu_{2l'+1}a_r) \\
&=& a_l(u_1u_2^{-1}u_3\cdots u_{2l'}^{-1}u_{2l'+1})a_r=a_lb_la_r. \end{eqnarray*}
\item[{\rm 4.}]  Alice computes $K:=a_l(b_la_rb_r)$. Bob also computes the shared key $(a_lb_la_r)b_r=K$.
\end{description} 

It is easy to show that this scheme is a further instance of the generalized AAG scheme for monoids (section \ref{AAGmon}).
Therefore one simply has to turn the sets $SK_A^{(r)}$ and $SK_B^{(l)}$ into monoids by introducing some ``forgetful'' operations as exercised, e.g., in the proof of \ref{grDH}. \\ 
In order to break this scheme an attacker obviously has to solve the following
\begin{list}{}{\setlength{\itemsep}{0cm} \setlength{\parsep}{0cm} }
\item[{\bf Base Problem:}]
\item[{\sc Input:}] Element pairs $(s_1,s'_1),\ldots ,(s_m,s'_m)\in G^2$ and $(t_1,t'_1),\ldots ,(t_n,t'_n)\in G^2$ with $s'_i=b_ls_ib_r$ $\forall 1\le i\le m$ and 
$t'_j=a_lt_ja_r$ $\forall 1\le j\le n$ for some (unknown) $a_l,b_r\in G$, $b_l\in SK_B^{(l)}$, $a_r\in SK_A^{(r)}$.  
\item[{\sc Objective:}] Find $K=a_lb_la_rb_r$.
\end{list}

A successful attack on Alice's secret key requires the solution of the following
\begin{list}{}{\setlength{\itemsep}{0cm} \setlength{\parsep}{0cm} }
\item[{\bf $n$-simDP} ($n$-Simultaneous Decomposition Problem):]
\item[{\sc Input:}]  Element pairs $(t_1,t'_1),\ldots ,(t_n,t'_n)\in G^2$ with $t'_j=a_lt_ja_r$ $\forall 1\le j\le n$ for some (unknown) $a_l\in G$, $a_r\in SK_A^{(r)}$.  
\item[{\sc Objective:}] Find elements $a'_l\in G$, $a'_r\in SK_A^{(r)}$ with $a'_lt_ja'_r=t'_j$ for all $j=1,\ldots ,n$.
\end{list}
A solution $(a'_l,a'_r)$ to this $n$-simDP satisfies the property $a'_lya'_r=a_lya_r$ for all $y\in SK_B^{(l)}$. \\ 
Analogeously, a successful attack on Bob's secret key requires the solution of the following
\begin{list}{}{\setlength{\itemsep}{0cm} \setlength{\parsep}{0cm} }
\item[{\bf $m$-simDP} ($m$-Simultaneous Decomposition Problem):]
\item[{\sc Input:}]  Element pairs $(s_1,s'_1),\ldots ,(s_m,s'_m)\in G^2$ with $s'_i=b_ls_ib_r$ $\forall 1\le i\le m$ for some (unknown) $b_l\in SK_B^{(l)}$, $b_r\in G$.  
\item[{\sc Objective:}] Find elements $b'_l\in SK_B^{(l)}$, $b'_r\in G$ with $b'_ls_ib'_r=s'_i$ for all $i=1,\ldots ,m$.
\end{list}
A solution $(b'_l,b'_r)$ to this $m$-simDP satisfies the property $b'_lxb'_r=b_lxb_r$ for all $x\in SK_A^{(r)}$. \\
Therefore, a solution to both problems provides the attacker with the shared secret, because
\[ (a'_lb'_la'_r)b'_r=(a_lb'_la_r)b'_r=a_l(b'_la_rb'_r)=a_l(b_la_rb_r)=K. \]
Here the first and the last equality hold, because $b'_l\in SK_B^{(l)}$ and $a_r\in SK_A^{(r)}$, respectively.
Alternatively, we can use equality chain
\[ a'_l(b'_la'_rb'_r)=a'_l(b_la'_rb_r)=(a'_lb_la'_r)b_r=(a_lb_la_r)b_r=K, \]
where here the first and the last equality hold, because $a'_r\in SK_A^{(r)}$ and $b_l\in SK_B^{(l)}$, respectively.
Further, the first equality chain shows us, that it is sufficient to find a solution $(a'_l,a'_r)\in G^2$ to the $n$-SDP and a solution $(b'_l,b'_r)\in SK_B^{(l)}\times G$ to the
$m$-simDP.
Analogously, the second equality chain shows us, that it is sufficient to find a solution $(a'_l,a'_r)\in G\times SK_A^{(r)}$ to the $n$-SDP and a solution 
$(b'_l,b'_r)\in G^2$ to the $m$-simDP. \\
Note that the knowledge of one secret key, e.g. Alice's key $(a_l,a_r)\in G\times SK_A^{(r)}$, is not sufficient for an attacker to obtain the shared
secret $K$, because he needs not only $a_r$ expressed in the generators of the group $G$, but rather an expression of the form
\[  a_r=r_1r_2^{-1}r_3r_4^{-1}\cdots r_{2l}^{-1}r_{2l+1} \quad {\rm with} \quad r_i\in \{s_1,\ldots ,s_m\}. \]

{\bf Remark.} An attacker might approach an $n$-simDP instance $\{ (t_i, t'_i) \}_{i\le n}$ by considering the ${ n \choose 2}$-ssCSP instance 
$\{ (t'_i(t'_j)^{-1}, t_it_j^{-1}) \mid 1 \le i \ne j \le n \}$ or the ${ n \choose 2}$-ssCSP instance 
$\{ (t_i^{-1}t_j, (t'_i)^{-1}t'_j) \mid 1 \le i \ne j \le n \}$  in order to solve for $a_l$ or $a_r$, respectively. For example, in the latter case, we have
\[ a_r^{-1}t_i^{-1}t_ja_r=(a_r^{-1}t_i^{-1}a_l^{-1})(a_lt_ja_r)=(t'_i)^{-1}t'_j. \] 
Therefore, either the simultaneous (subgroup)-CSP has to be hard in $G$, or, if the simCSP is (at least heuristically) approachable in $G$,
it is recommended that the sets $\{ t_i^{-1}t_j \mid 1 \le i \ne j \le n \}$ and $\{ t_it_j^{-1} \mid 1 \le i \ne j \le n \}$
have large centralizers. This may be ensured by if the set $\{ t_1, \ldots ,t_n \}$ itself has a large centralizer, an thus also $S_B$.s
Similarly $S_A$ should have a large centralizer. 

\subsection{Non-associative KEP based on simultaneous symmetric DP} \label{symDP} 

Here we consider the following specifications of the AAG scheme for magmas: \\
Let $k,l\in \mathbb{N}$ be $G=M=N=S_1=S_2$ be a group. The group multiplication symbol in $G$ will usually be omitted.  
The operations $\bullet _i,\circ _i$ ($i=1,2$) on $G$ are defined as in the previous subsection by
\[ x\bullet _1y=x\bullet _2y=x\circ _1y=x\circ _2y\equiv x\bullet y:=xy^{-1}x, \]
and the functions $\beta _1,\beta _2: G\times G\rightarrow G$ are defined by
\[ \beta _1(x,y)=x^kyx, \quad \beta _2(x,y)=xyx^l. \]
$\beta _i (x,\cdot )$ ($i=1, 2$) fulfills the homomorphy condition $(1)$ for all $x\in G$, because
\begin{eqnarray*} \beta _i(x,y_1)\bullet \beta _i(x,y_2) &=& (x^ky_1x^l)\bullet (x^ky_2x^l)= \\
(x^ky_1x^l)x^{-l}y_2^{-1}x^{-k}(x^ky_1x^l) &=& x^k(y_1y_2^{-1}y_1)x^l=\beta _i(x,y_1\bullet y_2), 
\end{eqnarray*} 
where either $k=1$ or $l=1$.  \\
Alice and Bob publicly assign sets $\{s_1,\ldots ,s_m\}, \{t_1,\ldots ,t_n\}\subset G$, respectively. 
The secret key spaces of Alice and Bob are the submagmas $S_A=\langle s_1,\ldots ,s_m\rangle _{\bullet }$ and $S_B=\langle t_1,\ldots ,t_n\rangle _{\bullet }$ 
of $(G,\bullet )$ generated by the publicly assigned elements. \\
The projections $\pi_1, \pi_2$ are the identity ${\rm id}_G$, and the functions $\gamma _1,\gamma _2: G\times G\rightarrow G$ are defined by 
\[  \gamma _1(x,y)=x^ky, \,\, \gamma _2(x,y)=yx^l. \]
These definitions satisfy condition $(3)$, which gives the shared key
\[ \gamma _1(a, \beta (b,\pi_1(a)))=\gamma _1(a,bab^l)=a^kbab^l=\gamma _2(b,a^kba)=\gamma _2(b,\beta (a,\pi_2(b))).  \]

Consider the simultaneous version of symmetrical decomposition problem (see \cite{CDW07}). 
\begin{list}{}{\setlength{\itemsep}{0cm} \setlength{\parsep}{0cm} }
\item[{\bf $n$-sim $(k,l)$-SDP} ($n$-simultaneous $(k,l)$-Symmetrical Decomposition Problem):]
\item[{\sc Input:}]  Integers $(k, l) \in \mathbb{Z}^2$ and element pairs 
$(t_1,t'_1),\ldots ,(t_n,t'_n)\in G^2$ with $t'_j=a^kt_ja^l$ $\forall 1\le j\le n$ for some (unknown) $a\in G$.  
\item[{\sc Objective:}] Find elements $a'\in G$ with $a'^kt_ja'^l=t'_j$ for all $j=1,\ldots ,n$.
\end{list}

We conclude that an attack on Alice's or Bob's private key has to master an $n$-sim $(k,1)$-SDP or an $m$-sim $(1,l)$-SDP, respectively. \\

{\bf Remark.} One may also consider a variant of that KEP where the integers $k, l$ are parts of Alice's and Bob's secrret key. In particular, set
$S_1=S_2=\mathbb{Z} \times G$, $\pi _1=\pi _2: (p, x) \mapsto x$, $\beta _1((k,a),y)=a^kya$, $\beta _2 ((l,b),y)=byb^l$, 
$\gamma _1((k,a),v)=a^kv$, and $\gamma _2((l,b),v)=vb^l$. Then an attack, e.g. on Alice's secret key, has provide $k\in \mathbb{Z}$ and $a\in G$
such that $a^kt_ja=t'_j$ for all $j$.

\section{Non-associative schemes for LD-systems} \label{LD}
\subsection{LD- and multi-LD-systems} \label{LDmult} 
\subsubsection{Definition}

\begin{defi} \hspace{-0.15cm}{\bf .} 
An {\it LD-system} $(S,*)$ is a set $S$ equipped with a binary operation $*$ on $S$ which satisfies the {\it left-selfdistributivity law} 
\[ x*(y*z)=(x*y)*(x*z) \quad \forall x,y,z\in S. \] 
\end{defi}

\begin{defi}\hspace{-0.15cm}{\bf .}  (Section X.3. in \cite{De00})
Let $I$ be an index set. A {\it multi-LD-system}
 $(S,(*_i)_{i\in I})$ is a set $S$ equipped with a family of binary operations $(*_i)_{i\in I}$ on $S$ such that 
\[ x*_i(y*_jz)=(x*_iy)*_j(x*_iz) \quad \forall x,y,z\in S \]
is satisfied for every $i,j$ in $I$. Especially, it holds for $i=j$, i.e., $(S,*_i)$ is an LD-system. If $|I|=2$ then we call $S$ a {\it bi-LD-system}. 
\end{defi}

A classical example for an LD-system is given by a group $G$ equipped with the conjugacy operation $x*y=x^{-1}yx$.
We also mention the Laver tables (Chapter X in \cite{De00}) as standard examples for finite monogenic LD-systems.
Many examples for LD-, bi-LD- and multi-LD-systems are given in Dehornoy's monography \cite{De00}.

\subsubsection{$f$-conjugacy} 

One may consider several generalizations of the conjugacy operation as candidates for natural LD-operations in groups.
Consider an Ansatz like $x*y=f(x^{-1})g(y)h(x)$ for some group endomorphisms $f,g,h$.

\begin{prop} \hspace{-0.15cm}{\bf .} \label{PropLDConj}
Let $G$ be a group, and $f,g,h \in End(G)$. Then the binary operation
$x*y=f(x^{-1}) \cdot g(y)\cdot h(x)$ yields an LD-structure on $G$ if and only if 
\begin{equation} \label{fConjLDeqs} fh=f, \quad gh=hg=hf, \quad fg=gf=f^2, \quad h^2=h. \end{equation}
\end{prop}

{\sc Proof}. A straightforward computation yields
\begin{eqnarray*}
\alpha * (\beta * \gamma )&=&f(\alpha ^{-1}) gf(\beta ^{-1}) g^2(\gamma ) gh(\beta ) h(\alpha ), \quad {\rm and} \\
(\alpha * \beta)*(\alpha *\gamma )&=& fh(\alpha ^{-1}) fg(\beta ^{-1}) f^2(\alpha ) gf(\alpha ^{-1}) g^2(\gamma ) gh(\alpha ) hf(\alpha ^{-1}) hg(\beta ) h^2(\alpha ).
\end{eqnarray*}
A comparison of both terms yields the assertion. \quad $\Box $ \\

The simplest solution of the system of equations (\ref{fConjLDeqs}) is $f=g$ and $h={\rm id}$. This leads to the following definition. 
\begin{defi} \hspace{-0.15cm}{\bf .} ({\sc LD- or $f$-conjugacy}) \label{fConj}
Let $G$ be a group, and $f\in End(G)$. An ordered pair $(u, v)\in G \times G$ is called
$f$-LD-conjugated or LD-conjugated, or simply $f$-conjugated, denoted by $u\stackrel{}{\longrightarrow }_{*_f}v$, if $\exists c\in G$ such that
$v=c*_f u=f(c^{-1}u)c$.
\end{defi}

{\bf Remark.} For any non-trivial endomorphism $f$, the relation $\stackrel{}{\longrightarrow }_{*_f}$ defines not an equivalence relation on $G$.
Even the relation $\stackrel{}{\longrightarrow }_{*}$ defined by $u\stackrel{}{\longrightarrow }_*v$ iff $\exists f \in Aut(G)$ s.t. $u\stackrel{}{\longrightarrow }_{*_f}v$ is not an equivalence relation. Indeed, transitivity requires the automorphisms (relation must be symmetric!) to be an idempotent endomorphism ($f^2=f$) which implies $f={\rm id}$. \\
Compare the notion of $f$-LD-conjugacy with the well known notion {\it $f$-twisted conjugacy} defined by $u \sim _f v$ (for $f\in Aut(G)$) iff  
$\exists c\in G$ s.t. $v=f(c^{-1})uc=:c*^{tw}_f u$, which yields indeed an equivalence relation.
On the other hand, the operation $*^{tw}=*^{tw}_f$ is not LD - rather it satisfies the following "near" LD-law:
\[ \alpha *^{tw} (\beta *^{tw} \gamma )=(\alpha *^{tw} \beta)*^{tw}(\alpha ^f *^{tw}\gamma ) \]
where $\alpha ^f$ is short for $f(\alpha )$.  \\
Anyway, it follows directly from the definitions that $u\stackrel{}{\longrightarrow }_*v$ if and only if $f(u) \sim _f v$, i.e., any $f$-LD conjugacy problem
reduces to a twisted conjugacy problem and vice versa. Here we have to extend the notion of twisted conjugacy from $f\in Aut(G)$ to all $f\in End(G)$.

\subsubsection{Shifted conjugacy}
Patrick Dehornoy introduced the following generalization of $f$-conjugacy, and he points out, that once the definition of shifted conjugacy is used, braids inevitably appear \cite{De00,De06}.

\begin{prop} \hspace{-0.15cm}{\bf .}  \label{ExI.3.20.} {\rm (Exercise I.3.20. in \cite{De00})}
Consider a group $G$, a homomorphism $f: G\rightarrow G$, and a fixed element $a\in G$. Then the binary operation
\[ x*y=x *_{f,a}y=f(x)^{-1} \cdot a \cdot f(y)\cdot x  \]
yields an LD-structure on $G$ if and only if $[a,f^2(x)]=1$ for all $x\in G$, and $a$ satisfies the relation $a f(a) a=f(a) a f(a)$. 
Hence the subgroup $H=\langle \{ f^n(a) \mid  n\in \mathbb{N} \} \rangle $ of $G$ is a homomorphic image of
the braid group
\[ B_{\infty }= \langle \{\sigma _i \}_{i\ge 1} \mid \sigma _i\sigma _j=\sigma _j\sigma _i \,\,{\rm for} \,\, |i-j|\ge 2, \,\, 
\sigma _i\sigma _j\sigma _i=\sigma _j\sigma _i\sigma _j \,\, {\rm for} \,\, |i-j|=1\rangle \] 
with infinitely many strands, i.e., up to an isomorphism, it is a quotient of $B_{\infty }$. 
\end{prop}

There exists a straightforward generalization of Proposition \ref{ExI.3.20.} for multi-LD-systems:

\begin{prop} \hspace{-0.15cm}{\bf .} \label{multiLD} Let $I$ be an index set. Consider a group $G$, a family of endomorphisms $(f_i)_{i\in I}$ of $G$, and a set of fixed elements $\{a_i\in G \mid i\in I\}$. 
Then $(G,(*_i)_{i\in I})$ with
\[ x*_iy= f_i(x^{-1})\cdot a_i \cdot f_i(y)\cdot x\]
is a multi-LD-system if and only if $f_i=f_j=:f$ for all $i\ne j$, $[a_i,f^2(x)]=1$ for all $x\in G$, $i\in I$, and $a_if(a_i)a_j=f(a_j)a_if(a_i)$ for all $i,j\in I$. 
\end{prop}

{\sc Proof}. A straightforward computation gives
\begin{eqnarray*} x*_i(y*_jz)&=&f_i(x^{-1})a_i [f_i(f_j(y^{-1})) f_i(a_j) f_i(f_j(z)) f_i(y)]x, \\
(x*_iy)*_j(x*_iz)&=& [f_j(x^{-1})  f_j(f_i(y^{-1}))  f_j(a_i^{-1})  f_j(f_i(x))]  a_j   [ f_j(f_i(x^{-1})) \cdot \\
 && f_j(a_i)  f_j(f_i(z))  f_j(x)][ f_i(x^{-1})  a_i f_i(y)  x].
\end{eqnarray*}
A comparison of both terms yields the assertion. \quad $\Box $ \\

Note that this proof also contains proofs of Proposition \ref{ExI.3.20.} (setting $|I|=1$) and of the following Corollary \ref{ShConj} (setting $G=B_{\infty }$, $I=\{1,2\}$, $s=\partial $, $*_1=*$, $*_2=\bar{*}$, $a_1=\sigma _1$ and $a_2=\sigma _1^{-1}$). \\

Consider the injective {\it shift endomorphism} $\partial : B_{\infty } \longrightarrow B_{\infty }$ defined by $\sigma _i \mapsto \sigma _{i+1}$ forall $i\ge 1$.
\begin{coro}  \hspace{-0.15cm}{\bf .}  \label{ShConj} {\sc (Shifted conjugacy}, {\rm Example X.3.5. in \cite{De00})} 
$B_{\infty }$ equipped with the {\rm shifted conjugacy} operations $*$, $\bar{*}$ defined by
\[ x*y=\partial x^{-1}\cdot \sigma _1 \cdot \partial y \cdot x, \quad \quad  x\, \bar{*}\, y=\partial x^{-1}\cdot \sigma _1^{-1} \cdot \partial y  \cdot x  \]
is a bi-LD-system. In particular, $(B_{\infty },*)$ is an LD-system.
\end{coro}

\subsubsection{Generalized shifted conjugacy in braid groups}

In the following we consider generalizations of the shifted conjugacy operations $*$ in $B_{\infty }$. Therefore we
set $f=\partial ^p$ for some $p\in \mathbb{N}$, and we choose $a_i\in B_{2p}$ for all $i\in I$ such that 
\begin{equation} a_i\partial ^p(a_i)a_j=\partial ^p(a_j)a_i\partial ^p(a_i) \quad \forall i,j\in I.  \end{equation}  
Since $a_i\in B_{2p}$, we have $[a_i,\partial ^{2p}(x)]=1$ for all $x\in B_{\infty }$.
Thus the conditions of Proposition \ref{multiLD} are fulfilled, and $x*_iy=x\partial ^p(y)a_i\partial ^p(x^{-1})$ defines an multi-LD-structure on $B_{\infty}$. 
For $|I|=1$, $p=1$ and $a=\sigma _1$, which implies $H=B_{\infty }$, we get Dehornoy's original definition of shifted conjugacy $*$. \\
It remains to give some natural solutions $\{a_i\in B_{2p} \mid i\in I \}$ of the equation set (1). Note that in case $|I|=1$ (notation: $a_1=a$), of course, every
endomorphism $f$ of $B_{\infty }$ with $f(\sigma _1) \in B_{2p}$ provides such solution $a=f(\sigma _1)$. 

\begin{defi} (Definition I.4.6. in \cite{De00}) Let, for $n\ge 2$, be $\delta _n=\sigma _{n-1}\cdots \sigma _2\sigma _1$. For $p, q \ge 1$, we set
\[ \tau _{p,q}=\delta _{p+1}\partial (\delta _{p+1})\cdots \partial ^{q-1}(\delta _{p+1}). \]
\end{defi}

Since $a=\tau _{p,p}^{\pm 1}\in B_{2p}$ fulfills $a\partial ^p(a)a=\partial ^p(a)a\partial ^p(a)$, it provides a lot of (multi)-LD-structures on $B_{\infty }$.  

\begin{prop} \label{abc} (a) The binary operation $x*_ay=\partial ^p(x^{-1})a\partial ^p(y)x$ with $a=a'\tau _{p,p}a''$ for some $a',a''\in B_p$ yields an LD-structure on $B_{\infty }$ 
if and only if $[a',a'']=1$. \\
(b) Let $I$ be an index set.
The binary operations $x*_iy=\partial ^p(x^{-1})a_i\partial ^p(y)x$ with $a_i=a'_i\tau _{p,p}a''_i$ for some $a'_i,a''_i\in B_p$ ($i\in I$) yields a multi-LD-structure on $B_{\infty }$ 
if and only if $[a_i',a_j']=[a_i',a_j'']=1$ for all $i,j\in I$. (Note that $a_i''$ and $a_j''$ needn't commute for $i\ne j$.) \\
(c) The binary operations $x*_iy=\partial ^p(x^{-1})a_i\partial ^p(y)x$ ($i=1,2$) with $a_1=a_1'\tau _{p,p}a_1''$, $a_2=a_2'\tau _{p,p}^{-1}a_2''$ for some $a_1',a_1'',a_2',a_2''\in B_p$ 
yields a bi-LD-structure on $B_{\infty }$ if and only if $[a_1',a_1'']=[a_2',a_2'']=[a_1',a_2'']=[a_2',a_1'']=[a_1',a_2']=1$.
(Note that $a_1''$ and $a_2''$ needn't commute.)  
\end{prop}

Another solution
We see that there exist infinitely many (multi)-LD-structures on $B_{\infty }$. Further examples are provided by Proposition \ref{split}, which, of course, admits a lot of variations and generalizations.

\begin{prop} \label{split} Let be $p,p_1,p_2\in \mathbb{N}$ with $p_1+p_2=p$.  The binary operation $x*_ay=\partial ^p(x^{-1})a\partial ^p(y)x$ with 
\[ a=a_1'\partial ^{p_1}(a_2')\partial ^{p_1}(\tau _{p_2,p})\tau _{p,p_1}^{-1}a_1''\partial ^{p_1}(a_2'')\] for some $a_1',a_1''\in B_{p_1}$, $a_2',a_2''\in B_{p_2}$ yields an LD-structure 
on $B_{\infty }$ if and only if $[a_1',a_1'']=[a_2',a_2'']=1$. 
\end{prop}

The proofs of Proposition \ref{abc} and \ref{split} are straightforward computations. The reader is recommended to draw some pictures.

\subsubsection{Yet another group-based LD-system} \label{YetLD}
Though we are sure that it must have been well known to experts, we haven't been able to find the following natural LD-operation for groups in the literature.
For a group $G$, $(G,\circ )$ is an LD-system with $x\circ y=xy^{-1}x$. \\
Note that, contrary to the conjugacy operation $*$, for this {\it "symmetric decomposition" or conjugacy operation} $\circ $, 
the corresponding relation $\stackrel{}{\longrightarrow }_{\circ }$ defined by $x\stackrel{}{\longrightarrow }_{\circ }y$ iff $\exists c\in G$ such that $y=c\circ x$) is not an equivalence relation. In particular, $\stackrel{}{\longrightarrow }_{\circ }$ is reflexive and symmetric, but not transitive. \\ 
One may consider several generalizations of this symmetric conjugacy operation $\circ $, as candidates for natural LD-operations in groups.
Consider an Ansatz like $x\circ y=f(x)g(y^{-1})h(x)$ for some group endomorphisms $f,g,h$.

\begin{prop} \hspace{-0.15cm}{\bf .} \label{PropLDsymmConj}
Let $G$ be a group, and $f,g,h \in End(G)$. Then the binary operation
$x\circ y=f(x) \cdot g(y^{-1})\cdot h(x)$ yields an LD-structure on $G$ if and only if 
\begin{equation} \label{fsymmConjLDeqs} f^2=f, \quad fh=gh=fg, \quad hg=gf=hf, \quad h^2=h. \end{equation}
\end{prop}

{\sc Proof}. A straightforward computation yields
\begin{eqnarray*}
\alpha \circ (\beta \circ \gamma )&=&f(\alpha ) gh(\beta ^{-1}) g^2(\gamma ) gf(\beta ^{-1}) h(\alpha ), \quad {\rm and} \\
(\alpha \circ \beta)\circ (\alpha \circ \gamma )&=& f^2(\alpha ) fg(\beta ^{-1}) fh(\alpha ) gh(\alpha ^{-1}) g^2(\gamma ) gf(\alpha ^{-1}) hf(\alpha ) hg(\beta ^{-1}) h^2(\alpha ).
\end{eqnarray*}
A comparison of both terms yields the assertion. \quad $\Box $ \\

Except for $f^2=f=g=h=h^2$, the simplest solutions of the system of equations (\ref{fsymmConjLDeqs}) are $f^2=f=g$ and $h={\rm id}$, or
$f={\rm id}$ and $g=h=h^2$.  
\begin{coro} \hspace{-0.15cm}{\bf .} ({\sc LD- or $f$-symmetric conjugacy}) \label{fsymC}
Let $G$ be a group, and $f\in End(G)$ an endomorphism that is also a projector ($f^2=f$).
Then $(G, \circ _f)$ and $(G, \circ _f^{\rm rev})$, defined by $x\circ _f y=f(xy^{-1})x$ and $x \circ _f^{\rm rev} y=xf(y^{-1}x)$, are LD-systems.
\end{coro}

\begin{prop} \hspace{-0.15cm}{\bf .} \label{distrOversymmConj}
Let $G$ be a group, and $f, g \in End(G)$. \\
(i) Then the binary operations $\circ _f$ and $*_f$ (and $*^{\rm rev}_f$), defined by 
$x\circ _f y=f(x) \cdot g(y^{-1})\cdot h(x)$ and $x*_f y=f(x^{-1} \cdot y)\cdot h(x)$ ($x* ^{\rm rev}_f y=x \cdot f(y \cdot x^{-1})$),
are distributive over $\circ $. In particular $*$ ($*^{\rm rev}$) is distributive over $\circ $. In short, the following equations hold.
\[   x*_f(y\circ z)=(x*_fy)\circ (x*_fz), \quad x\circ _f(y\circ z)=(x\circ _fy)\circ (\circ _fz) \forall x,y,z\in G.\]
(ii) The operations $\circ _f$ and $*_f$ ($*^{\rm rev}_f$) are distributive over $\circ _g$ if and only if $f=gf=fg$.
\end{prop}

\subsection{Non-associative AAG $f$-commutator KEP} \label{AAGfCommKEP}

Now we consider the most natural special case of our general AAG scheme for magmas (see section \ref{AAGmagma}). Let be $M=N=S$. This implies that the functions $\beta _i, \gamma _i$, for $i=1,2$, induce further binary operations on $M$. In particular, we introduce the notation $x*_iy=\beta _i(x,y)$.
Now, the homomorphy condition $(1)$ (in section \ref{AAGmagma}) reads as
\begin{eqnarray*} x*_1(y \bullet _2y')&=&(x*_1y)\circ _2(x*_1y') \quad {\rm and} \\
 x*_2(y\bullet _1y')&=&(x*_2y)\circ _1(x*_2y').
\end{eqnarray*} 
If $\bullet _i=\circ _i$ holds for $i=1,2$, then $M$ fulfills two distributive laws. And if additionally $\circ _2=\circ _1=*_1=*_2=:*$, then $(M,*)$ is an LD-system. \\
We observe that LD-systems occur in a very natural special case of the general AAG scheme for magmas. Nevertheless, this does not imply that we get by that construction KEP's for all LD-systems. Indeed, in order to obtain a shared key, we have to specify the projections $\pi _1$ and binary operations $\gamma _i$ which themselves depend on the specification of the LD-operation $*$. In the following we set $\pi _i={\rm id}_M$ for $i=1, 2$. \\
Now, we establish a (non-associative) AAG-KEP for groups with $f$-conjugacy as LD-operation.
Let $M=G$ be a group, $f\in End(G)$, then $(G,+)$ with $*=*_f$ (see Def. \ref{fConj}) is an LD-system according to Proposition \ref{PropLDConj}.

\begin{defi} \hspace{-0.15cm}{\bf .} ({\sc $f$-commutator}) \label{fComm}
Let $G$ be a group, and $f\in End(G)$. The {\rm $f$-commutator} of an ordered pair $(u, v)\in G \times G$ is defined by
\[ [u,v]_f:=u^{-1} f(v^{-1}) f(u)v. \]
\end{defi}

The {\it AAG $f$-commutator KEP} is given by the following further specifications of the general AAG scheme for magmas (section \ref{AAGmagma}).
\[ \gamma _1(u,v)=u^{-1}v, \quad \quad \gamma _2(u,v)=v^{-1}u. \]

Now, Alice and Bob perform the following protocol steps.

\begin{description}
\item[{\rm 1.}] Alice generates her secret key $a$ in the public submagma $S_1=\langle s_1, \cdots , s_m \rangle _*$ of $(G,*)$, and Bob chooses his secret key $b\in S_2=\langle t_1, \cdots , t_n \rangle _*$.
\item[{\rm 2.}] Alice computes the elements $a*t_1, \ldots ,a*t_n\in G$, and sends them to Bob.
Analogously Bob computes the elements $b*s_1, \ldots ,b*s_m\in G$, and sends them to Alice. 
\item[{\rm 3.}] Alice, knowing $a=T_*(r_1, \ldots , r_k)$ with $r_i\in \{s_1,\ldots ,s_m\}$, computes from Bob's public key
\[ T_*(b*r_1, \ldots , b*r_k)=b*T_*(r_1, \ldots , r_k))=b*a=f(b^{-1}a)b. \]
And Bob, knowing $b=T'_*(u_1, \ldots , u_{k'})$ with $u_j\in \{t_1,\ldots ,t_n\}$, computes from Alice's public key
\[ T'_*(a*u_1, \ldots , a*u_{k'})=a*T'_*(u_1, \ldots , u_{k'})=a*b=f(a^{-1}b)a. \]
\item[{\rm 4.}] Alice computes $K:=\gamma _1(a,b*a)=a^{-1}(b*a)=a^{-1}f(b^{-1}a)b=[a,b]_f$.
Bob gets the shared key by $\gamma _2(b,a*b)=(a*b)^{-1}b=(f(a^{-1}b)a)^{-1}b\stackrel{(3)}{=}K$.
\end{description}

In order to break this scheme an attacker obviously has to solve the following base problem.
\begin{list}{}{\setlength{\itemsep}{0cm} \setlength{\parsep}{0cm} }
\item[{\bf $f$-AAGP} ({\it $f$-Commutator AAG-Problem}):]
Let $(G,*)$ be a group with  $\alpha * \beta =f(\alpha ^{-1}\beta )\alpha $ for some $f\in End(G)$. 
Furthermore, let $A=\langle a_1,\ldots ,a_k\rangle _*$ and $B=\langle b_1, \ldots ,b_m\rangle _*$ be two f.g. submagmas of $(G,*)$.
\item[{\sc Input:}] $\{ (a_i,y*a_i) \in G^2 | i=1,\ldots ,k\} \cup \{ (b_j,x*b_j) \in G^2 | j=1,\ldots ,m\}$ with $x \in A$ 
and $y \in B$.
\item[{\sc Objective:}] Find the $f$-commutator $[x,y]_f:=x^{-1}f(y^{-1}x)y$.
\end{list}

But a successful attack on Bob's secret key requires at least the solution of the following
\begin{list}{}{\setlength{\itemsep}{0cm} \setlength{\parsep}{0cm} }
\item[{\bf $m$-sim $f$-CSP} ($m$-Simultaneous $f$-Conjugacy Search Problem):]
\item[{\sc Input:}]  Pairs $(s_1,s'_1),\ldots ,(s_m,s'_m)\in G^2$ with $s'_i=b*s_i=f(b^{-1}s_i)b$ $\forall 1\le i\le m$ 
for some (unknown) $b\in G$.  
\item[{\sc Objective:}] Find an element $b'\in G$ with $f(b'^{-1}s_i)b'=f(b^{-1}s_i) b'$ for all $i=1,\ldots ,m$.
\end{list}

Even if one solves that problem, one might have not found Bob's original secret $b$. This raises the question of how rigid solutions to the
simultaneous $f$-CSP are. A vague indication for some kind of rigidity is the fact that $f(b'b^{-1})$ and $b'b^{-1}$ are conjugated with every $f(s_i)$
($1\le i \le m$) being a valid conjugator. \\
Anyway, even if an attacker finds Bob's original key $b$, then she still faces the following problem.

\begin{list}{}{\setlength{\itemsep}{0cm} \setlength{\parsep}{0cm} }
\item[{\bf $*_f$-MSP} ($*_f$-submagma Membership Search Problem):]
\item[{\sc Input:}] $b, t_1,\ldots ,t_n\in G$.
\item[{\sc Objective:}] Find an expression of $b$ as a tree-word in the submagma $\langle t_1,\ldots ,t_n \rangle _{*_f}$ (notation $b=T_{*_f}(u_1,\ldots ,u_k)$ for $u_i \in \{ t_j \} _{j\le n}$), if it exists.
\end{list}

Another approach is to attack (additionally to Bob's secret key) also Alice's key, i.e., to solve for the $n$-simultaneous $f$-CSP-instance $\{(t_j,t'_j)\}_{j\le n}$ with $t'_j=f(a^{-1}t_j)a$.
An oracle to that problem provides an element $a'\in G$ such that $t'_j=f(a'^{-1}t_j)a'$ for all $j$. Then the attacker hopes that
computation of the $f$-commutator $[a',b']_f=:K'$ might give her the shared key $K=[a,b]_f$. \\
Though the $f$-CSP seems to be particulary interesting for non-invertible endomorphism $f\in End(G)$, here we compare $K'$ with $K$ for the simplest case
where $f\in Inn(G)$, i.e., there exists an element $p\in G$ s.t. $f(x)=p^{-1}xp$. Then it is easy to show that $b'b^{-1}=:c_1$ lies in $\bigcap _i C_G(s_ip)$,
and $a'a^{-1}=c_2\in \bigcap _j C_G(t_jp)$.
A straightforward computation gives
\[ K'=a'^{-1}p^{-1}b'^{-1}a'pb'=a^{-1}c_2^{-1}p^{-1}b^{-1}c_1^{-1}c_2apc_1b . \]
We conclude that $K'=K$ if $[c_1,c_2]=[c_1,ap]=[c_2,bp]=1$. But, in general, we have $C_G(ap) \ne \bigcap _i C_G(s_ip)$ and $C_G(bp) \ne \bigcap _j C_G(t_jp)$. Therefore, even in the case of $f\in Inn(G)$, we can't hope to reduce the $f$-AAGP to a simultaneous subgroup CSP, as we have done it for the classical AAGP in Proposition \ref{BasePs}. \\
Nevertheless, as in the remark at the end of section \ref{relNCsch}, one may  approach an $n$-sim $f$-CSP instance $\{ (t_i, t'_i) \}_{i\le n}$ by considering the ${ n \choose 2}$-simCSP instance 
$\{ (t_i^{-1}t_j, (t'_i)^{-1}t'_j) \mid 1 \le i \ne j \le n \}$  in order to solve for $a$. Indeed, here we have
\[ a^{-1}t_i^{-1}t_ja=(a^{-1}f(t_i^{-1}a))(f(a^{-1}t_j)a_r)=(t'_i)^{-1}t'_j. \] 
Therefore, either the simultaneous CSP has to be hard in $G$, or, if the simCSP is (at least heuristically) approachable in $G$,
it is recommended that the sets $\{ t_i^{-1}t_j \mid 1 \le i \ne j \le n \}$ and $\{ t_it_j^{-1} \mid 1 \le i \ne j \le n \}$
have large centralizers. This may be ensured by if the set $\{ t_1, \ldots ,t_n \}$ itself has a large centralizer.
Similarly $\{ s_1, \ldots , s_m \}$ should have a large centralizer.  

\subsubsection{An example in pure braid groups} \label{Pn}

Here we a concrete suggestion for the group $G$ and the endomorphism $f\in End(G)$. Let $G$ be the $n$-strand pure braid group $P_n$.
For some small integer $d\ge 1$, consider the epimorphism $\eta _d: P_n \longrightarrow P_{n-d}$ given by 'pulling out' (or erasing) the last $d$ strands, i.e. the strands $n-d+1, \ldots , n$. Recall the shift map $\partial $, and note that $\partial ^d (P_{n-d}) \le P_n$. Now, we define the endomorphism $f: P_n \longrightarrow P_n$ by the composition $f=\partial ^d \circ \eta $.  

\subsection{Non-associative AAG shifted commutator KEP in braid groups} \label{AAGshCommKEP}

Here we establish a (non-associative) AAG-KEP for braid  groups with shifted conjugacy as LD-operation.
Recall from Corollary \ref{ShConj} that the braid group $(B_{\infty },*,\bar{*})$ forms a bi-LD-system. Also recall the definition of shift endomorphism 
$\partial $.

\begin{defi} \hspace{-0.15cm}{\bf .} {(\sc shifted commutator)} \label{shComm}
The {\rm shifted commutator} of an ordered pair $(u, v)\in B_{\infty } ^2$ is defined by
\[ [u,v]_{\rm sh}:=u^{-1} \partial(v^{-1}) \sigma _1 \partial(u)v. \]
\end{defi}

The {\it AAG shifted commutator KEP} for the bi-LD-system $(B_{\infty },*,\bar{*})$ is given by the following further specifications of the general AAG scheme for magmas (section \ref{AAGmagma}). \\
Set $M=N=S=B_{\infty }$, $\pi _i={\rm id}_M$,  $\beta _i(x,y)=:x*_iy$, $\bullet _i=\circ _i =*_i$ for $i=1, 2$, and 
\begin{eqnarray*}
x*_1y=x\, \bar{*}\, y=\partial (x^{-1})\sigma _1^{-1}\partial (y)x, && 
x*_2y=x*y=            \partial (x^{-1})\sigma _1     \partial (y)x, \quad {\rm and} \\
\gamma _1(u,v)=u^{-1}v, &&  \gamma _2(u,v)=v^{-1}u. 
\end{eqnarray*}

Now, Alice and Bob perform the following protocol steps.

\begin{description}
\item[{\rm 1.}] Alice generates her secret key $a$ in the public submagma $S_1=\langle s_1, \cdots , s_m \rangle _{\bar {*}}$ of $(B_{\infty },*,\bar{*})$, and Bob chooses his secret key $b\in S_2=\langle t_1, \cdots , t_n \rangle _*$.
\item[{\rm 2.}] Alice computes the elements $a\bar {*}t_1, \ldots ,a\bar {*}t_n\in G$, and sends them to Bob.
Analogously Bob computes the elements $b*s_1, \ldots ,b*s_m\in G$, and sends them to Alice. 
\item[{\rm 3.}] Alice, knowing $a=T_{\bar {*}}(r_1, \ldots , r_k)$ with $r_i\in \{s_1,\ldots ,s_m\}$, computes from Bob's public key
\[ T_{\bar {*}}(b*r_1, \ldots , b*r_k)=b*T_{\bar {*}}(r_1, \ldots , r_k)=b*a=\partial (b^{-1}) \sigma _1 \partial (a)b. \]
And Bob, knowing $b=T'_*(u_1, \ldots , u_{k'})$ with $u_j\in \{t_1,\ldots ,t_n\}$, computes from Alice's public key
\[ T'_*(a\bar {*}u_1, \ldots , a\bar {*}u_{k'})=a\bar {*}T'_*(u_1, \ldots , u_{k'})=a\bar {*}b=\partial (a^{-1})\sigma _1^{-1} \partial (b)a. \]
\item[{\rm 4.}] Alice computes $K:=\gamma _1(a,b*a)=a^{-1}(b*a)=a^{-1}\partial (b^{-1}) \sigma _1 \partial (a)b=[a,b]_{\rm sh}$.
Bob gets the shared key by $\gamma _2(b,a\bar {*}b)=(a\bar {*}b)^{-1}b=(\partial(a^{-1}) \sigma _1^{-1} \partial (b)a)^{-1}b\stackrel{(3)}{=}K$.
\end{description}

In order to break this scheme an attacker obviously has to solve the following base problem.
\begin{list}{}{\setlength{\itemsep}{0cm} \setlength{\parsep}{0cm} }
\item[{\bf sh-AAGP} (shifted Commutator AAG-Problem):]
Consider the bi-LD-system $(B_{\infty },*,\bar{*})$. 
Let $A=\langle a_1,\ldots ,a_k\rangle _{\bar{*}}$ and $B=\langle b_1, \ldots ,b_m\rangle $ be two f.g. submagmas of $(B_{\infty },*,\bar{*})$.
\item[{\sc Input:}] $\{ (a_i,y*a_i) \in G^2 | i=1,\ldots ,k\} \cup \{ (b_j,x\bar{*}b_j) \in G^2 | j=1,\ldots ,m\}$ with $x \in A$ 
and $y \in B$.
\item[{\sc Objective:}] Find the shifted commutator $[x,y]_{\rm sh}:=x^{-1}\partial (y^{-1}) \sigma _1 \partial (x)y$.
\end{list}

But a successful attack on Bob's secret key requires at least the solution of the following
\begin{list}{}{\setlength{\itemsep}{0cm} \setlength{\parsep}{0cm} }
\item[{\bf $m$-sim sh-CSP} ($m$-simultaneous shifted Conjugacy Search Problem):]
\item[{\sc Input:}]  Pairs $(s_1,s'_1),\ldots ,(s_m,s'_m)\in G^2$ with $s'_i=b*s_i=\partial (b^{-1}) \sigma _1 \partial (s_i)b$ $\forall 1\le i\le m$ 
for some (unknown) $b\in G$.  
\item[{\sc Objective:}] Find an element $b'\in G$ with $\partial (b'^{-1}) \sigma _1 \partial (s_i)b'=\partial (b^{-1}) \sigma _1 \partial (s_i) b'$ for all $i=1,\ldots ,m$.
\end{list}

As in the case of $f$-conjugacy, one may argue that finding $b$ is not sufficient, since the attacker still faces a submagma MSP for $(B_{\infty },*,\bar{*})$. 
Furthermore, as for $*_f$, one may show that solving two simultaneous sh-CSP's (for Alice's and Bob's private keys) does in general not reduce the sh-AAGP
to a simultaneous subgroup CSP, as for the classical AAGP. \\

{\bf Remark.} Note that we actually do not need a bi-LD-system, like $(B_{\infty },*,\bar{*})$, in order to build a AAG shifted commutator KEP.
Indeed, two LD-operations, namely $x*y=\partial (x^{-1}) \sigma _1 \partial (y)x$ and its reverse $x*^{\rm rev}y= x \partial (y) \sigma _1 \partial (x^{-1})$, suffice. Here $(B_{\infty }, * ,  *^{\rm rev})$ is not a bi-LD-system. \\
Alice and Bob choose $a\in \langle s_1,\ldots , s_m \rangle _*$ and $b\in \langle t_1,\ldots , t_n \rangle _{*^{\rm rev}}$,
and send $\{ a^{-1}*^{\rm rev} t_j \}_{j \le n}$ and  $\{ b^{-1}* s_i \}_{i \le m}$, respectively. Then they may compute
\[ K_A=a^{-1}(b^{-1}*a)= a^{-1} \partial (b) \sigma _1 \partial (a) b^{-1}=[a, b^{-1}]_{\rm sh} =(a^{-1}*^{\rm rev} b)b^{-1}=K_B. \]
Analogeously, one may build an AAG $f$-commutator KEP using $*_f$ and its reverse operation. 

{\bf Non-simultaneity.} Analogeous to the remarks in sections \ref{relNCsch} and \ref{AAGfCommKEP}, an attacker
 might approach an $m$-sim shCP instance $\{ (s_i, s'_i=b*s_i) \}_{i\le m}$ by considering the ${ m \choose 2}$-simCSP instance  
$\{ (\partial (s_i^{-1}s_j), (s'_i)^{-1}s'_j) \mid 1 \le i \ne j \le m \}$  in order to solve for $b$. Indeed, here we have
\[ b^{-1}\partial(s_i^{-1}s_j)b=(b^{-1}\partial (s_i^{-1})\sigma _1^{-1}\partial (b))(\partial (b^{-1}) \sigma _1 \partial (s_j)b)=(s'_i)^{-1}s'_j. \] 
Therefore, either it is recommended that the set $\{ s_i^{-1}s_j \mid 1 \le i \ne j \le m \}$ (and analogeously $\{ t_i^{-1}t_j \mid 1 \le i \ne j \le n \}$)
has large centralizer. This may be ensured by if the sets $\{ s_1, \ldots ,s_m \}$ and $\{ t_1, \ldots ,t_n \}$ itself have a large centralizer. \\
Another strategy is to abandon simultaneity, i.e, to consider the critical case $m=n=1$. Note that only for shifted conjugacy (and its generalizations)
we have opportunity to abandon simultaneity because only here the submagmas $\langle s \rangle _*$, $\langle s \rangle _{\bar *}$ generated by one element are nontrivial. This is not the case for $f$-conjugacy or the LD-operation $\circ $ from section \ref{YetLD}.

{\bf Generalized shifted conjugacy.} It is straightforward to construct non-associative KEP's using generalized shifted conjugacy operations.
We leave this to the reader.

\section{Generalizations, further work and open problems}  \label{Open}
\subsection{AAG-schemes over non-associative and non-commutative algebras}

It is possible to generalize the AAG-KEP for magmas from section \ref{AAGmagma} in several ways.
One generalization is very simple - just replace the magmas $(M,\bullet _1, \bullet _2)$ and $(N,\circ _1, \circ _2)$ by
$(M, \{\bullet _{1,i}\}_{i\in I_1}, \{\bullet _{2,i}\}_{i\in I_2})$ and $(N, \{\circ _{1,i}\}_{i\in I_1}, \{\circ _{2,i}\}_{i\in I_2})$ for some index sets $I_1, I_2$, i.e. we introduce further binary operations. In particular, in the special case given by $M=N=S_1=S_2$ and $\pi _1=\pi _2= {\rm id}_M$, Alice chooses her secret key $a$ as an element from the submagma $\langle s_1, \ldots , s_m \rangle _{\{ \bullet _{1,i} \}_{i\in I_1}}$. \\
To describe an element of such a submagma it is not sufficient to know the planar rooted binary tree $T$ (providing the bracket structure) and the leaf elements
$r_1,\ldots , r_k \in \{ s_1, \ldots , s_m \}$, but we also need to assign binary operations (from the set $\{ \bullet _{1,i} \}_{i\in I_1}$) to the internal nodes of the tree $T$. \\
For example, \\
In the following we write $T_{\{ \bullet _{1,i} \}_{i\in I_1}}$, and we assume that $T$ is then a planar rooted binary tree accompanied with such an assignment of its internal nodes. \\
Here we have to modify condition (1) from section \ref{AAGmagma} in the obvious way: \\

(1') $\beta _1 (x,\cdot ): (M, \bullet _{2,i}) \rightarrow (N, \circ _{2,i})$ is for all $x\in S_1$, $i\in I_2$ a magma homomorphism, i.e. 
\[ \forall x\in S_1, \,\, y, y'\in M, \,\, i\in I_2: \quad \beta _1(x,y\bullet _{2,i} y')=\beta _1(x,y)\circ _{2,i}\beta _1(x,y'). \]
Also $\beta _2 (x,\cdot ): (M, \bullet _{1,i}) \rightarrow (N, \circ _{1,i})$ is for all $x\in S_2$, $i\in I_1$ a magma morphism, i.e. 
\[ \forall x\in S_2, \,\, y, y'\in M, \,\, i\in I_1: \quad \beta _2(x,y\bullet _{1,i} y')=\beta _2(x,y)\circ _{1,i}\beta _2(x,y'). \]

If $\beta _1, \beta _2$ are defined by a binary operation from a bi- or multi-LD-system, then condtion (1') is satisfied by construction.
Now one may build KEP's with this obvious modification. One example is the AAG shifted commutator KEP for the bi-LD-system $(B_{\infty },*,\bar{*})$.
Indeed, there Alice and Bob may have choosen their secret keys from  $\langle s_1, \cdots , s_m \rangle _{*, \bar {*}}$ of $(B_{\infty },*,\bar{*})$ and $\langle t_1, \cdots , t_n \rangle _{*,\bar{*}}$, respectively.  \\
Recall that bi- and multi-LD-systems fulfill more homomorphic properties (i.e. distributive laws) than is necessary to build a KEP.
As an example, consider the group ring $\mathbb{Z}G$. Recall that $(G, *_f)$ is an LD-system for any $f \in End(G)$. By construction, $(\mathbb{Z}G, *_f, +)$ is a non-commutative and non-associative algebra. It is straightforward to build a non-associative KEP over $\mathbb{Z}G$ analogous to the non-associative AAG $f$-commutator KEP. The only modification is that we choose the secret keys $a\in \langle s_1, \cdots , s_m \rangle _{*_f, +}$ and $b \in \langle t_1, \cdots , t_n \rangle _{*_f, +}$ for $s_1, \cdots , s_m, t_1, \cdots , t_n \in \mathbb{Z}G$. \\
Analogoulsly, it is straightforward to build a non-associative KEP over the non-associative bialgebra $(\mathbb{Z}B_{\infty }, *, \bar{*}, +)$. \\
Furthermore, one could consider non-commutative (but associative) special cases of these KEP's over non-associative algebras, if one restricts the secret
keys $a, b$ (or more precisely the projectio n $\pi _1(a)$, $\pi _2(b)$) to $\langle s_1, \cdots , s_m \rangle _+$ and $b \in \langle t_1, \cdots , t_n \rangle _+$, respectively.

\subsection{Open problems and further work}
  
\begin{itemize}
\item The AAG-KEP for magmas (see section \ref{AAGmagma}) describes a general framework for building non-associative key establishment protocols.
Our main examples are provided by LD-operations ($f$-conjugacy in groups and shifted conjugacy in braid groups). Recall also the systems based on
(simultaneous) symmetric DP employing the non-associative operation given by $x\bullet y=xy^{-1}x$.  \\
Find other interesting instances of the general AAG-KEP for magmas (see section \ref{AAGmagma}).
\item Find other platform groups for the non-associative AAG $f$-commutator KEP (see section \ref{AAGfCommKEP}). \\
In particular, solve the (simultaneous) $f$-conjugacy problem in pure braid groups for the endomorphism $f$ described in section \ref{Pn}. 
\item How rigid are the solutions to the $f$-conjugacy problem in pure braid groups and the shifted conjugacy problem in braid groups?
Note that, contrary to the $f$-conjugacy problem in pure braid groups, there exists a solution to the shifted conjugacy problem in braid groups \cite{KLT09}.
\item Investigate heuristic attacks, especially length-based attacks \cite{HT02, GK+05}, on the submagma MSP for non-associative LD-operations $*$ in braid groups.
Of particular interest is here the non-simultaneous case $m=1$ which emerges only for non-associative operations.
I.e. consider the submagma MSP for the submagma $\langle s_1 \rangle _*$ generated by only one element.
\item Recall the important special case of the AAG-KEP for magmas where $S=M=N$ is an LD-system. Depending on the LD-operation $*$, we constructed for some
instances non-associative KEP's by specifying the functions $\gamma _i$ ($i=1, 2$). It would be nice to have non-associative KEP's for all LD-systems. \\
Such non-associative KEP's for all LD-systems, bi-LD- and multi-LD-systems (in general: sets with distributive operations) have been constructed - see our forthcoming paper \cite{KaT12}. There we have to go even a step beyond the general AAG-KEP for magmas, and we introduce a small asymmetry in the non-associative AAG protocol. Indeed, we consider the systems and instances given in \cite{KaT12} as more practical and interesting than the one given in this paper.
Since the KEP's given in \cite{KaT12} work for all multi-LD-systems, they deploy two further advantages. \\
(1) We can consider encryption functions using iterated $*$-multiplication from the left. In order to obtain the secret key an attacker has to solve then an
iterated $f$- or shifted conjugacy problem. \\
(2) For a given (partial) multi-LD-system $(M, \{ *_i \} _{i \in I})$ it turns out that even the used operations $*_i$ can be hidden, i.e., they are part of the secret key.
\item Develop other primitives like signature and authentication schemes in non-associative cryptography. \\
Here we concentrated on KEP's which are usually the hardest to construct. Note that, using hash functions, it is easy to build public key encryption schemes from KEP's.
\item For infinite groups, like braid groups, there are limitations on the depths of the trees describing a submagma element . Consider for example $f$-conjugacy in an infinte group $G$ where $f\in End(G)$ satisfies $|f(x)|\le |x|$ for all $x\in G$.
Denote by $|\cdot |=|\cdot |_X$ the word length over some given generating set $X$ of $G$. We conclude that
\[ |x*_fy|\le |f(x^{-1})|+|f(y)|+|x| \le	 2|x|+|y| \le 3\max \{ |x|, |y| \}.  \] 
Now, consider the following two extreme cases of trees with $k$ leaves defining the bracket structure of a magma element in $\langle s_1, \ldots , s_m \rangle _*$
($*=*_f$). The {\it left comb} ($r_j \in \{ s_i \}_{i\le m}$ for all $j=1, \ldots k$) 
\[ LC(r_1, \ldots , r_k):= r_1 *(r_2 * (r_3* \cdots r_{k-2}*(r_{k-1}*r_k) \cdots )), \]
and the {\it right comb}
\[ RC(r_1, \ldots , r_k):= (( \cdots (r_1 *r_2) * r_3* \cdots *r_{k-2}) *r_{k-1}) *r_k. \]
If $|s_i|\le l_0$ for all $i=1,\ldots , m$, then one may show by induction that
\[ |LC(r_1, \ldots , r_k)|\le (2k-1)l_0, \quad |RC(r_1, \ldots , r_k)| \le (2^k-1)l_0. \]
I.e. we can only prove an exponential (in $k$) upper bound on the word length of a magma element of tree depth $k-1$. 
But for left combs we have a linear upper bound. In practice, one may consider as keys either only elements of small tree depth, or we choose such elements
whose bracket structure defining trees have a small "distance" from a left comb. \\ 
Define a proper notion of "distance" of planar rooted binary trees, and investigate how the word length growth for trees with "small distance" from the left comb $LC$. Determine a method how such trees can be generated efficiently.
\item Recently B. Tsaban developed a deterministic polynomial time attack on the AAG commutator KEP in linear groups \cite{Ts12} which also applies to several other non-commutative schemes. 
In short, Tsaban's {\it linear centralizer attack} exploits the fact that in classical AAG-KEP  the shared key is the commutator
$K=a^{-1}b^{-1}ab$. So, if we find solutions (up to centralizer elements) inside the centralizer of the centralizer of, say $S_A$,
then these centralizer elements cancel and we recover $K$, even if these solutions were only in the linear matrix group in which we embed our linear group.
But for KEP's with shared key $K=a_lb_la_rb_r$, or  $K$ being an $f$-commutator in groups or a shifted commutator in braid groups,
these centralizer elements would not cancel. Therefore, we conclude that, in its present state the linear centralizer attack does not apply to most of
the non-associative schemes presented in this paper. \\
Can the linear centralizer attack be improved to make it work against these KEP's?
\end{itemize}

{\it E-mail address:} {\tt arkadius.kalka@rub.de} \\

SCHOOL OF MATHEMATICS AND PHYSICS, UNIVERSITY OF QUEENSLAND, BRISBANE, AUSTRALIA
\end{document}